\documentclass[12pt]{article}
\usepackage{epsfig}
\usepackage{bm}
\usepackage{amsmath}
\usepackage{amssymb}
\usepackage{graphics}
\usepackage[english]{babel}
\begin{document}


\vspace{1cm}
\title{\textbf{QCD analysis of Lambda hyperon production in DIS target-fragmentation region}}
\author{
\vspace{0.5cm}\\
\textbf{Federico Alberto Ceccopieri$^{(a,b)}$\footnote{Email address : federico.ceccopieri@hotmail.it}}
\ and 
\textbf{Davide Mancusi$^{(a,c)}$\footnote{Email address : davide.mancusi@cea.fr}}
\vspace{0.7cm}\\
\textit{$^{(a)}$IFPA, Universit\'e de Li\`ege,}\\
\textit{All\'ee du 6 ao\^ut, B\^at B5a, 4000 Li\`ege, Belgium}
\vspace{0.4cm}\\
\textit{$^{(b)}$Universit\'e Libre de Bruxelles,}\\
\textit{ Boulevard du Triomphe, 1050 Bruxelles, Belgium}
\vspace{0.4cm}\\
\textit{$^{(c)}$CEA, Centre de Saclay, IRFU/Service de Physique Nucl\'{e}aire,}\\
\textit{F-91191 Gif-sur-Yvette, France} }
\date{}


\maketitle
\thispagestyle{empty}
\vspace{0.5cm}
\begin{center}
\large
Abstract
\vspace{0.2cm}\\
\end{center}
\normalsize
We consider Lambda-hyperon production in the target-fragmentation region of semi-inclusive 
deep-inelastic scattering within the framework of fracture functions. 
We present a first attempt to determine the flavour and energy dependences 
of these non-perturbative distributions through a simultaneous QCD-based fit to available 
neutral- and charged-current semi-inclusive-DIS cross sections.
Predictions based on the resulting nucleon-to-Lambda fracture functions
are in good agreement with data and observables not included in the regression. The
successful prediction of
the $Q^2$ dependence of the Lambda multiplicity notably represents the first validation
of the perturbative framework implied by fracture functions.


\newpage
\section{Introduction}

It has been known for a long time~\cite{Basile} that in hadronic collisions the 
longitudinal momentum spectrum of particles produced in target fragmentation   
crucially depends on the difference of the valence-parton composition of the initial- and final-state particles. 
In particular, only an initial-state particle whose valence-quark flavour content 
is almost or totally conserved in the scattering 
can be a leading particle in the final state, \textsl{i.e.} carry a substantial fraction 
of the incoming projectile energy. Counting rules connect the differences 
in the valence-parton composition  of initial- and final- state particles with the shape 
of longitudinal momentum spectrum of the latter. 
Leading particles in the final state are typically characterised by large longitudinal momentum fractions
and very small transverse momenta with respect to the collision axis, 
a typical regime dominated by soft QCD dynamics where standard perturbative techniques 
cannot be applied.
 
The leading particle effect manifests itself in reactions involving at least one hadron in the initial state. 
Quite interestingly, it also appears in processes which involve point-like probes,
such as Semi-Inclusive Deep Inelastic Scattering (SIDIS).
At variance with the hadronic collisions discussed above, such process involves a large momentum transfer. Therefore, thanks to the factorisation theorem,  
the short-distance cross section can be calculated by using 
perturbation theory, while soft QCD effects are effectively encoded 
by universal parton distributions and fragmentation functions.
The properties of these non-perturbative objects
are generally extracted from the so called current-fragmentation region, 
\textsl{i.e.} the phase-space region in which the struck parton hadronises.
 
The target-fragmentation region is instead sensitive to
the hadronisation properties of the coloured spectator system which results
from the removal of one parton from the incident nucleon by the virtual probe.
For these reasons, the description of particle production in this particular region 
of phase space through standard perturbative 
calculations based on parton distributions and fragmentation functions will fail when compared to data. 
The description can be improved only with the introduction of new non-perturbative distributions which encode
these peculiar aspects of soft QCD dynamics. 

This was early realised in Ref.~\cite{trentadue_veneziano} 
where the authors introduced the concept of fracture functions. 
These distributions simultaneously encode information on the
interacting parton and on the fragmentation of the spectator system.
Although intrinsically of non-perturbative nature, the scale dependence of such distributions 
can be calculated within perturbative QCD~\cite{trentadue_veneziano}.
Fracture functions obey in fact
DGLAP~\cite{DGLAP} inhomogeneous evolution equations
which result from the structure of collinear singularities 
in the target-fragmentation region~\cite{trentadue_veneziano,graudenz}.
Moreover, a dedicated factorisation theorem~\cite{factorization_soft,factorization_coll} 
guarantees that fracture functions 
are universal distributions, at least in the context of SIDIS. 

The phenomenology which makes use of all these concepts is still
confined to the study of hard diffraction in DIS. 
Within this framework, in fact, no analysis has been attempted 
for particles other than proton in the final state. 
It is well known, however, that Lambda hyperon
production in SIDIS is mainly concentrated in the target-fragmentation 
region since Lambdas show a significant leading particle effect.
  
In this paper we describe how a variety of Lambda leptoproduction cross sections
can be simultaneously described within the fracture-function approach, if 
these non-perturbative distributions are modelled at some low scale and their 
free parameters determined by a fit to the available data.  
The rather scarce Lambda leptoproduction data in general do not allow to directly verify 
the leading-twist nature of this type of processes, which is implicitly assumed in the fracture-function framework, 
nor allow to test the scale dependence embodied by their specific evolution equations.
In this respect, both the formalism and the model presented in this paper 
require more experimental information for a conclusive validation.
We believe, however, that a quantitative tool that is able to reproduce many aspects of the existing data 
may further stimulate both theoretical and experimental activity. As a by-product,
the model will give us the first insights on the flavour and energy dependences 
of the fragmentation properties of the spectator system into Lambda hyperons.

The paper is organized as follows. In Section~\ref{TFR_CS} we briefly recall 
SIDIS cross sections and fracture-function properties.
In Section~\ref{Data_observables} we discuss specific features of the data sets 
and observables used in the analysis. In Section~\ref{FFmodel} we describe 
a simple model for Lambda fracture functions and in Section~\ref{Fit}
we provide and discuss the results of our fit. In Section~\ref{predictions}
we compare the model predictions with data and observables not used in the fit.
In Section~\ref{Conclusions} we summarise our results. 

\section{Semi-Inclusive DIS in the target-fragmentation region}
\label{TFR_CS}

The deep inelastic scattering cross section of a lepton $l$ off a proton $p$ with four-momenta 
$k$ and $P$, respectively, is described in terms of the lepton variables:
\begin{equation}
\label{variables0}
x_B=\frac{Q^2}{2 P\cdot q} , \;\;\; y=\frac{P\cdot q}{P\cdot k}=\frac{Q^2}{s_h x_B} ,\;\;\; Q^2=-q^2, 
\end{equation}
where $k'$ and $q=k-k'$ are the outgoing lepton and virtual photon four-momenta, respectively, 
$s_h=(P+k)^2$ 
is the centre of mass energy squared and
$W^2=s_h y (1-x_B)+m_p^2$ is the invariant mass squared of the final state, with $m_p^2$ being the proton mass. 
The additional invariant~\cite{graudenz}
\begin{equation}
\label{variables1}
z_h=  \frac{P \cdot h}{P \cdot q} =  \frac{E_h }{E_P (1-x_B)} \frac{1-\cos \theta_h}{2} ,
\end{equation}
is often used to specify the kinematics of final state hadron with four-momentum $h$,
where $E_h$ and $\theta_h$ are the detected hadron energy and angle respectively  
defined in the virtual photon-proton centre of mass frame.
The variable $z_h$ is however not adequate to describe target fragmentation, 
since both soft hadron production ($E_h \simeq 0$) and hadron production 
in the target-remnant direction ($\theta_h \simeq 0$) both yield a vanishing 
value of $z_h$.
We therefore consider cross sections differential
either in the scaled hadron energy variable $z$ or $\zeta$~\cite{graudenz}  
\begin{equation}
\label{variables2}
z = \frac{\zeta}{1-x_B}, \;\;  \zeta=\frac{E_h}{E_P},
\end{equation}
again defined in the $\gamma^*p$ centre of mass frame. 
It follows from eq.~(\ref{variables2}) that 
final-state hadrons are detected with a fraction
$z\in[0,1]$ of the spectator energy $E_p(1-x_B)$.
In the following we will analyse data presented in term 
of Feynman's variable 
\begin{equation}
\label{variables3}
x_F = \pm \Big(z^2-\frac{4 m_T^2}{W^2}\Big)^{\frac{1}{2}}\,.
\end{equation}
We adopt the convention that, in the $\gamma^*p$ frame, negative values of $x_F$
correspond to final state hadrons moving parallel to incoming
proton direction.
In eq.~(\ref{variables3}) we have introduced the hadron transverse mass, $m_T^2=m_h^2+p_{h,\perp}^2$, defined  
in terms of its transverse momentum and mass squared. 

In the quark-parton model, the neutral-current semi-inclusive DIS cross section for producing an unpolarised Lambda off a proton in the target-fragmentation region reads~\cite{graudenz}
\begin{equation}
\label{LOeP}
\frac{d^3 \sigma^{lp \rightarrow l\Lambda X}}{dx_B dQ^2 d\zeta}=\frac{2\pi\alpha_{em}^2}{Q^4} J Y_+ \sum_{q=u,d,s} e_q^2 \Big[M_q^{\Lambda/p}(x_B,Q^2, \zeta)+
M_{\bar{q}}^{\Lambda/p}(x_B,Q^2, \zeta)\Big]\,,
\end{equation}
where $Y_+=1+(1-y)^2$. The cross section has been re-expressed for later convenience 
in term of the $\zeta$ variable, and the jacobian $J=\zeta[(1-x_B)|x_F|]^{-1}$ has been explicitely indicated~\cite{Mulders}.
The latter reduces to unity in the high-energy limit and it is therefore often omitted in the literature.
The neutrino- and anti-neutrino-induced charged-current semi-inclusive cross sections
 read respectively
\begin{equation}
\label{LOnuP}
\frac{d^3 \sigma^{\nu p \rightarrow \mu^- \Lambda X}}{dx_B dQ^2 d\zeta}=\frac{2\pi\alpha_{em}^2}{Q^4}
J 8 \eta_W  \Big[ 2( M_{d}^{\Lambda/p} + M_{s}^{\Lambda/p} ) + 
2(1-y)^2 M_{\bar{u}}^{\Lambda/p} \Big]\,,
\end{equation}
and 
\begin{eqnarray}
\label{LOnubarP}
\frac{d^3 \sigma^{\bar{\nu}p \rightarrow \mu^+ \Lambda X}}{dx dQ^2 d\zeta}=\frac{2\pi\alpha_{em}^2}{Q^4}
J 8 \eta_W \Big[ 2 ( M_{\bar{d}}^{\Lambda/p} + M_{\bar{s}}^{\Lambda/p}) + 
2(1-y)^2 M_{u}^{\Lambda/p} \Big]\,,
\end{eqnarray}
where the dependencies of $M_{i}$ appearing in eq.~(\ref{LOnuP}) and eq.~(\ref{LOnubarP}) 
are to be understood as in eq.~(\ref{LOeP}). The factor $\eta_W$ is defined in terms 
of the Fermi constant $G_F$, the $W$-boson mass $M_W^2$ and the electromagnetic coupling constant $\alpha_{em}$ as~\cite{PDG}
\begin{equation}
\eta_W=\frac{1}{2}\Bigg( \frac{G_F M_W^2}{4 \pi \alpha_{em}} 
\frac{Q^2}{Q^2+M_W^2}\Bigg)^2\,.
\end{equation}
As appropriate for a lowest-order calculation, we have assumed a vanishing
longitudinal structure-function contribution in all formulas. We have further
neglected charm quarks contribution. 

In eqs.~(\ref{LOeP},~\ref{LOnuP},~\ref{LOnubarP}) the production of unpolarised Lambdas
in the remnant direction is described by fracture functions 
$M_i^{\Lambda/p}(x,\zeta,Q^2)$~\cite{trentadue_veneziano}.
These distributions express the probability to find a parton of flavour $i$ with fractional 
momentum $x_B$ and virtuality $Q^2$ conditional to the detection of a target Lambda 
with a fraction $\zeta$ of the incoming proton energy.
The scale dependence of fracture functions is given by the following evolution equations
\cite{trentadue_veneziano}
\begin{eqnarray}
\label{FFevo}
& &\frac{\partial M_{i}^{\Lambda/p}\left( x_B,\zeta, \mu^2\right)}{\partial \log \mu^2} = \frac{\alpha_s(\mu^2)}{2\pi} \int_{\frac{x_B}{1-\zeta}}^1 \frac{du}{u} \, P_{i}^{j} (u) 
  \, M_{j}^{\Lambda/p}\left( \frac{x_B}{u},\zeta, \mu^2\right) +\nonumber \\
&&+ \frac{\alpha_s(\mu^2)}{2\pi}\int^{\frac{x_B}{x_B+\zeta}}_{x_B} \frac{du}{x_B(1-u)} 
\,\widehat{P}_{i}^{j,l} (u) f_{j/p} \left(\frac{x_B}{u},\mu^2\right)\, D^{\Lambda}_l 
\left(\frac{\zeta u}{x_B (1-u)},\mu^2\right)\,,
\end{eqnarray}    
where  $P_i^j(u)$ and $\widehat{P}_i^{j,l} (u)$ are the regularised~\cite{DGLAP}
and real~\cite{veneko} Altarelli-Parisi splitting functions, respectively.
Eq.~(\ref{FFevo}) describes both processes  which contribute
to Lambda production in the target-remnant direction. 
The homogeneous term on the right hand side of eq.~(\ref{FFevo}) takes into account 
the effects of collinear parton radiation   
by the struck parton $i$ while the Lambda originates from the fracture function itself. 
The inhomogeneous one instead takes into account the possibility that the detected Lambda
results from the fragmentation of the radiated parton $l$, emitted collinearly to the incoming parton $j$. 
This term in fact is a convolution of parton distributions $f_{j/p}(x_B,Q^2)$ and 
fragmentation functions $D^{\Lambda}_l(z,Q^2)$.

Given the explorative purpose of this analysis, we use leading-order formulas for semi-inclusive 
cross sections and consistently solve fracture-function evolution equations 
at leading logarithmic accuracy. 
We note, however, that the full formalism is available at next-to-leading-order accuracy 
for both unpolarised~\cite{graudenz} and polarised~\cite{npb1,npb2} processes.

\section{Data sets and Observables}
\label{Data_observables}

\begin{table}[t]
\begin{center}
\begin{tabular}{cccccc} \hline \hline
Reaction & $\langle E_i \rangle$ & $\langle W^2 \rangle$ & 
$\langle Q^2 \rangle$ & $\langle x_B \rangle$ & $\Lambda$ rates \\
type & (Ge$V$)   & (Ge$V^2$)    &(Ge$V^2$)& & (\%)      \\  \hline
$\nu p       $      \cite{Chang} &  50.0  & - & - & - & $7.0\pm 1.2$  \\ 
$\nu n       $      \cite{Chang} &  50.0  & - & - & - & $7.0\pm 0.8$  \\  
$\nu p$             \cite{WA21}  &  42    & 34.7 & 8.7 &  0.2 & $5.2\pm0.3$ \\
$\bar{\nu} p $      \cite{WA21}  &  38.5  & 20.4 & 5.2 &  0.2 & $5.7\pm0.4$ \\ 	 
$\mu p$             \cite{EMC}   &  280   & 130  &  12 & 0.11 &    -        \\
$\mu D_2$           \cite{EMC}   &  280   & 130  &  12 & 0.11 &    -        \\
$\mu D_2$           \cite{E665}  &  490   & 292  & 8.6 & 0.036& $7.8\pm1.6$ \\ \hline
\end{tabular}
\caption{\small{Data sets used in the present analysis. 
$\langle E_i \rangle$ is the average energy of the incoming lepton.
Average kinematics and production rates for the various data set, when available, are indicated.}}
\label{dataset}
\end{center}
\end{table}
The data used in the present analysis come from a variety of fixed-target 
experiments. We include neutrino and anti-neutrino SIDIS data
which are crucial in providing minimal quark-flavour discrimination.  
In particular, the stringent cuts ($x_B >$ 0.2 and $W>4$ GeV) in data from Ref.~\cite{Chang} enhance the sensitivity to valence-quark fracture functions;
data from Ref.~\cite{WA21} are expected to be an admixture of valence and sea contributions and therefore constrain the  
relative normalisations of the respective fracture functions.
We further include in the fit neutral-current SIDIS data at higher beam energy presented in Refs.~\cite{EMC,E665} in order
to provide the necessary information about the energy dependence of the 
cross sections and therefore on the $x_B$-dependence of fracture functions.

Some care should be used in the selection of the targets. Since charged-current
cross sections are significantly
lower than neutral-current ones, many experiments have used nuclear targets. 
The first consequence is that nuclear corrections to fracture functions might be accounted for. 
Even more important for the purposes of this analysis, the fragmentation process itself might be affected by the nuclear medium, both in the current~\cite{Hermes_xenon} and in the target~\cite{SKAT} fragmentation 
region. In particular, it has been recently reported in Ref.~\cite{SKAT} that strange-particle yields in neutrino-nuclei interactions are enhanced in the target 
region, probably due primary particles re-interaction with the nuclear medium.
Since the particle yields show a mild power dependence on the atomic number of the target,  
we include in the present analysis only proton- and deuteron-target data. 

We would like to mention two effects which might affect the absolute normalisations of the various data sets.
The first one is related to the definition of fracture functions. 
Many analyses quoted in Tab.~(\ref{dataset}) have estimated the contributions to the 
Lambda yield coming from the decay of higher-mass resonances. 
When not otherwise stated in the original publications, 
we interpret the published data as referring to an inclusive Lambda sample, 
that is the sum of promptly produced Lambdas and Lambdas coming from the decay of higher-mass resonances, 
corrected for unseen decay modes (a typical example is the $\Sigma^0 \rightarrow \Lambda \gamma$ 
decay mode discussed in Ref.~\cite{EMC}). The decay of higher-mass resonances in fact happens 
at time scales much larger than the ones typical of perturbative processes and their effects will be effectively incorporated in fracture functions.

The second issue is  related to the contamination of the Lambda yield by secondary Lambdas produced by the re-interaction of primary pions with detector material.
This effect has been intensively studied in Ref.~\cite{NOMAD} and estimated to contribute up to 20\% to 
the Lambda yield. It is unknown to us to which extent this correction has been properly estimated 
and applied to all data sets.

We close this Section discussing the choice and the reconstruction of the observable to be used in the fit. By definition, cross sections differential in the Lambda fractional energy  
are insensitive to the phase space region in which the latter has been produced. 
In this variable, the current- and target-fragmentation contributions
overlap and the extraction of the latter therefore crucially depends on the precision with which we describe current fragmentation with available Lambda fragmentation functions.  
In order to overcome this problem, we consider differential cross sections in the Feynman variable $x_F$ 
which offer, to lowest order, a kinematical  
separation of the two contributions. The use of such a variable is however not free 
from additional issues: Lambda-mass effects introduced via eq.~(\ref{variables3}) 
may be sizeable, as suggested by the values of the averaged hadronic final-state invariant mass $\langle W^2 \rangle$ quoted in Tab.~(\ref{dataset}). 
Such effects are however not compatible with the pQCD factorisation theorem. In
the present analysis, Lambda-mass effects are therefore applied a posteriori to
the Lambda leptoproduction cross sections $\sigma^{\Lambda}$, as described in Ref.~\cite{AKK}. The value in each $x_F$-bin is calculated as follows
\begin{equation}
\frac{1}{\sigma_{\mbox{\tiny{DIS}}}} \frac{\Delta \sigma_i^{\Lambda}}{\Delta x_F^i} = 
\frac{1}{\sigma_{\mbox{\tiny{DIS}}}} 
\frac{1}{\Delta x_F^i} \int dE_l \Phi(E_l) \int_{\Omega} dx_B d Q^2 \int^{1-x_B}_0 d\zeta \frac{d^3 \sigma^{\Lambda}(E_l)}{dx_B d Q^2 d\zeta}\, \Theta^i(x_F)\,,
\label{xF_corrected}
\end{equation}
where the index $i$ labels the $i$-th bin and the bin-size is specified by $\Delta x_F^i=x_{F}^{i+1}-x_{F}^{i}$, 
with $x_{F}^{i}$ representing the experimental bin-edges.
Mass corrections are enforced with the kinematical constraint $\Theta^i(x_F)=\theta(x_F-x_{F}^{i}) \,\theta(x_{F}^{i+1}-x_F)$,
with $x_F$ calculated via eq.~(\ref{variables3}). 
The label $\Omega$ stands for the set of cuts which define the DIS selection of
a given data set. 
The resulting differential cross sections are then normalised with respect to
the inclusive DIS cross section
\begin{equation}
\sigma_{\mbox{\tiny{DIS}}} = \int dE_l \Phi(E_l) \int_{\Omega} dx_B d Q^2 \frac{d^2 
\sigma^{\Lambda}(E_l)}{dx_B d Q^2}\,, 
\label{sigma_DIS}
\end{equation}
calculated with parton-distribution functions of Ref.~\cite{GRVproton}. Both
cross sections are integrated over
the lepton flux factor $\Phi(E_l)$ expressed in units of GeV$^{-1}$. For monochromatic electron and muon beams of energy $E_{l,0}$, the latter simply reduces 
to $\delta(E_l-E_{l,0})$. For neutrino and anti-neutrino 
beams we use the flux-factor parametrisations extracted by dedicated analyses~\cite{nuflux}.
We finally note that mass-corrected distributions are derived by using eq.~(\ref{variables2}) 
and eq.~(\ref{variables3}) and therefore require the knowledge of the Lambda transverse momentum. 
From the very precise data of Ref.~\cite{NOMAD} we know that the Lambda $p_t$-spectrum 
is dominated by $p_t$ values much smaller than its mass.
We therefore approximate in eq.~(\ref{variables3}) the transverse mass $m_T^2$ with Lambda mass $m_\Lambda^2$ whose 
value is taken to be $m_\Lambda=1115.683$ MeV~\cite{PDG}. 

\section{Modelling Lambda Fracture Functions}
\label{FFmodel}

The fracture formalism relies on the assumption that leading particle production 
in the target-fragmentation region is a leading-twist process. The latter has been 
strikingly confirmed by experimental observation of hard diffraction at HERA~\cite{ZEUSdiff,H1diff}. 
In the present context, however, the limited amount of data is often presented
as $Q^2$-integrated single-differential distributions, a fact which prevents any conclusion to be drawn. 
We only note that such hypothesis might be indirectly supported by the moderate value 
of the average $Q^2$ for the semi-inclusive reactions quoted in Tab.~(\ref{dataset}) 
and by the mild increase of the 
Lambda average multiplicity $\langle n(\Lambda) \rangle$ as a function $Q^2$, as seen 
in data~\cite{NOMAD}. 

Although the scale dependence of fracture functions is predicted by perturbative
QCD, these distributions still need to be modelled at some low scale and evolved
to scales relevant to the experiments; the resulting free parameters controlling
their input distributions can then be constrained by a fit to data.  These
distributions, however, depend upon the $x_B$ and $\zeta$ variables and on the
interacting parton flavour and therefore are expected to contain a large number
of free parameters.  In this respect, it would be possible to use as input the
parton distributions and the fragmentation functions of Regge-based models, such
as the quark-gluon string model (QGSM) of Ref.~\cite{QGSM}. This would reduce
the number of free parameters; in the present analysis, however, we have decided
to use realistic parton distributions determined from fits to DIS experiments
\cite{GRVproton} and a completely free input parametrisation of the
fragmentation functions in order to guarantee sufficient generality and
flexibility. We will compare the QGSM and our best-fit parametrisations at the
end of Section~\ref{Fit}.

Since the hard scattering process occurs on time scales much shorter 
than spectator-fragmentation ones, we assume that, at an arbitrary low scale $Q_0^2$, 
fracture functions factorise into a product of ordinary parton distributions $f_{i/p}(x_B,Q_0^2)$ and what we address as spectator-fragmentation 
functions $\widetilde{D}^{\Lambda/p}_i(z)$
\begin{equation}
\label{inputFF}
(1-x_B) \, M_{i}^{\Lambda/p}\left( x_B, \zeta ,Q^2_0 \right) =
M_{i}^{\Lambda/p}(x_B, z ,Q^2_0) = f_{i/p}(x_B,Q_0^2)  \widetilde{D}^{\Lambda/p}_i(z)\,,\; i=q,\bar{q},g\,.
\end{equation}
We take advantage of the sea-valence decomposition offered by parton 
distributions of Ref.~\cite{GRVproton} to further decompose the valence parton contributions as
\begin{equation}
\label{ic1}
M_{q=u,d}^{\Lambda/p}(x_B,z,Q_0^2) = q_{val}(x_B,Q_0^2) \widetilde{D}^{\Lambda/p}_{q_{val}}(z)+q_{sea}(x_B,Q_0^2) \widetilde{D}^{\Lambda/p}_{q_{sea}}(z)\,.
\end{equation}
For the non-perturbative spectator-fragmentation function we choose a simple functional 
form of the type
\begin{equation}
\label{zinput}
\widetilde{D}^{\Lambda/p}_{i}(z) =  \overline{N_i} z^{\alpha_i} \, (1-z)^{\beta_i}\,.
\end{equation}
In order to minimise correlations between parameters, the normalisation coefficients in eq.~(\ref{zinput}) are defined as follows
\begin{equation}
\overline{N_i}=N_i \Big[ \int_0^1 dz \, z^{\alpha_i} \, (1-z)^{\beta_i}\Big]^{-1},\;\;\; \alpha_i,\beta_i>-1 \;,
\label{renorm_Ni}
\end{equation}
and $N_i$ are then used as free parameters in the fit. 
The inclusion of deuteron-target data in the fit requires the knowledge of neutron-to-Lambda
fracture functions. As a first approximation we assume $u$-$d$ isospin symmetry
\begin{eqnarray}
M_{d}^{\Lambda/n}(x_B,z,Q^2)=M_{u}^{\Lambda/p}(x_B,z,Q^2),\nonumber\\ 
M_{u}^{\Lambda/n}(x_B,z,Q^2)=M_{d}^{\Lambda/p}(x_B,z,Q^2)\,.
\label{isospin}
\end{eqnarray}
Limiting ourselves to the discussion of the valence region and indicating in parenthesis the flavour structure
of the spectator system, we note that eqs.~(\ref{isospin}) implie
\begin{equation}
\widetilde{D}^{\Lambda/p}_{u(ud)}=\widetilde{D}^{\Lambda/n}_{d(ud)}  \;\;\; \mbox{and} \;\;\;  \widetilde{D}^{\Lambda/p}_{d(uu)}=\widetilde{D}^{\Lambda/n}_{u(dd)}.
\label{isospin2}
\end{equation}
 
Given the input distributions in eqs.~(\ref{inputFF},\ref{ic1}), set at an initial scale $Q_0^2=0.5$ Ge$V^2$, 
we numerically solve the fracture-functions evolution equations in eq.~(\ref{FFevo}) 
by using a finite-difference method in $x_B$-space in slices of $\zeta$. We use 
the leading-order proton parton distribution of Ref.~\cite{GRVproton} and $\Lambda$ 
fragmentation functions of Refs.~\cite{AKK} extracted taking into account target-hadron mass effects.
For consistency, we follow the original evolution scheme of Ref.~\cite{GRVproton}.
We evolve light-quarks fracture functions at leading-logarithmic accuracy  
within a fixed flavour-number scheme. Heavy-quarks effects are included in the running 
of the strong coupling at the heavy-quark mass thresholds and $\Lambda_{\mbox{\tiny{QCD}}}^{(n_f)}$ values
are taken from Ref.~\cite{GRVproton}.

As a final remark to this section we would like to discuss the impact of the 
inhomogeneous term in eq.~(\ref{FFevo}) on the observables entering the fit.
As a first step we have checked that, in the current region where the leading order semi-inclusive 
cross sections are proportional to the product of parton distributions and fragmentation functions,
a reasonable description of $\mu p$ data of Ref.~\cite{EMC} can be obtained both in shape and in normalisation
by using parton distributions and fragmentation functions of Refs.~\cite{GRVproton} and Ref.~\cite{AKK}, respectively. We then use fragmentation functions of Ref.~\cite{AKK}
in the calculation of the inhomogenous term appering in eq.~(\ref{FFevo}). 
At the observable level such term accumulates, as expected, at small $|x_F|$ as a result of the $z$ shape 
of Lambda fragmentation functions. Moreover it contributes only at the percent level to the $x_F$ distributions 
since it describes Lambda production from the fragmentation of radiated partons with transverse
momentum greater than the minimum momentum transfer involved in the process, that is $Q_0^2$.
As such it only contributes to the small radiative tail of the Lambda $p_t$-spectrum. 

The radiative contributions generated by the
inhomogeneous term and the universality of fragmentation functions could be studied in
detail in processes that do not show any leading particle effect, for
example the production of anti-Lambdas~\cite{EMC} or light
mesons~\cite{clas} off protons, or in target-like Lambda production
in a perturbative regime (\textit{i.e.} with sufficiently large transverse momentum,
$p_t\ge$ 1 GeV).

\section{Fitting procedure}
\label{Fit}

\begin{table}[t]
\begin{center}
\begin{tabular}{ccc} \hline \hline
Reaction type      \hspace{0.4cm}  & \hspace{0.4cm} partial $\chi^2$ 
\hspace{0.4cm} &  $\#$ fitted points \\ \hline
$\nu p       $      \cite{Chang}  &  4.77    & 5 \\ 
$\nu n       $      \cite{Chang}  &  3.25    & 5 \\ 
$\nu p$             \cite{WA21}   &  6.36    & 8 \\
$\bar{\nu} p $      \cite{WA21}   &  9.08    & 8 \\ 	 
$\mu p$             \cite{EMC}    &  9.90    & 8 \\
$\mu D_2$           \cite{EMC}    & 10.58    & 9 \\
$\mu D_2$           \cite{E665}   &  0.20    & 3 \\ \hline
\end{tabular}
\caption{\small{Partial $\chi^2$ contributions and number of points in the fit for each data set.}}
\label{Best-chi2}
\end{center}
\end{table}
\begin{table}[t]
\begin{center}
\begin{tabular}{c|ccc} \hline \hline
$\widetilde{D}^{\Lambda/p}_i$ &  $N_i$ & $\alpha_i$ & $\beta_i$ \\ \hline
$u_{val}$ & 0.046 $\pm$ 0.006 & 2.82 $\pm$ 1.19     & 0.39 $\pm$ 0.33 \\  
$d_{val}$ & 0.027 $\pm$ 0.006 & $=\alpha_{u_{val}}$ & 1.28 $\pm$ 0.51  \\
$q_{sea}$ & 0.078 $\pm$ 0.010 &        0            & 1.84 $\pm$ 0.63   \\ \hline
\end{tabular}  
\caption{\small{Best-fit values according to eq.~(\ref{zinput}).}}
\label{bestfit}
\end{center}  
\end{table}
Each of the assumed Lambda spectator-fragmentation functions in eq.~(\ref{zinput}) contains three free 
parameters (one normalisation and two exponents). Our fitting strategy is the following: we first tentatively assume
as many different fragmentation functions as possible types of struck partons
(valence or sea $u$, valence or sea $d$, other sea quarks, gluon).
The observable $1/\sigma_{\mbox{\tiny{DIS}}} \, d\sigma^{\Lambda}/dx_F$ is then
reconstructed via eq.~(\ref{xF_corrected}) and the best-fit parameter values are
determined using the
\texttt{MINUIT}~\cite{minuit} program. 
We assume that the uncertainties on the cross sections combine statistical and systematic errors so that
we use the simplest version of the $\chi^2$-function as a merit function,
although this neglects correlations between data points (the number of the
latter is rather limited and amounts to 46). We then study the eigenvalues and
the eigenvectors of the fit covariance matrix to identify any parameters that are
badly constrained by the fit. We fix them by making some assumptions about
their values and we repeat the fit with the remaining parameters.

With the available data, for example, the fit can not constrain distinct sea-quark
fracture functions. We therefore assume a common spectator-fragmentation function for all of them, 
$\widetilde{D}_{q_{sea}}^{\Lambda/p}$.
Moreover the fit is found to be insensitive to the choice of the gluon spectator 
fragmentation function. This is not unexpected since the gluon is coupled to electroweak probes only through 
higher orders.
For this reason, we fix the gluon spectator-fragmentation functions to be equal to the sea one,
$\widetilde{D}_{g}^{\Lambda/p}=\widetilde{D}_{q_{sea}}^{\Lambda/p}$, 
reducing the number of free parameters to nine.
In such a fit the smallest eigenvalues of the Hessian matrix correspond to eigenvectors
whose largest components are associated with the parameters $\alpha_i$. 
The poor determination of such parameters can be partly associated to mass effects
in the reconstruction of the observable in eq.~(\ref{xF_corrected}) via eq.~(\ref{variables3}).
The parameter $\alpha_{q_{sea}}$ is compatible with zero within errors so we fix
it to this value. Furthermore, the parameters $\alpha_{u_{val}}$ and
$\alpha_{d_{val}}$ are equal to each other
within errors, so we assume $\alpha_{u_{val}}=\alpha_{d_{val}}$, reducing
the number of free parameters to seven. 
Fixing some of the parameters to a definite value is indeed an 
arbitrary procedure. We can motivate this choice only a posteriori by noting 
that, when these parameters are optimsed by the fit, the value
of the $\chi^2$ function is only marginally reduced.
The best seven-parameter fit yields a $\chi^2/d.o.f.=44.14/(46-7)= 1.13$. 
The partial $\chi^2$ 
and the number of points included in the fit for each data set are displayed in Tab.~(\ref{Best-chi2}).
The results for the best-fit parameters are reported in Tab.~(\ref{bestfit}) along with parameter 
errors as calculated by the \texttt{MINUIT} routine \texttt{HESSE}, which
assumes a parabolic behaviour of the $\chi^2$ function in parameter space around the minimum.

\begin{figure}[t]
\begin{center}
\includegraphics[width=8cm,height=6cm,angle=0]{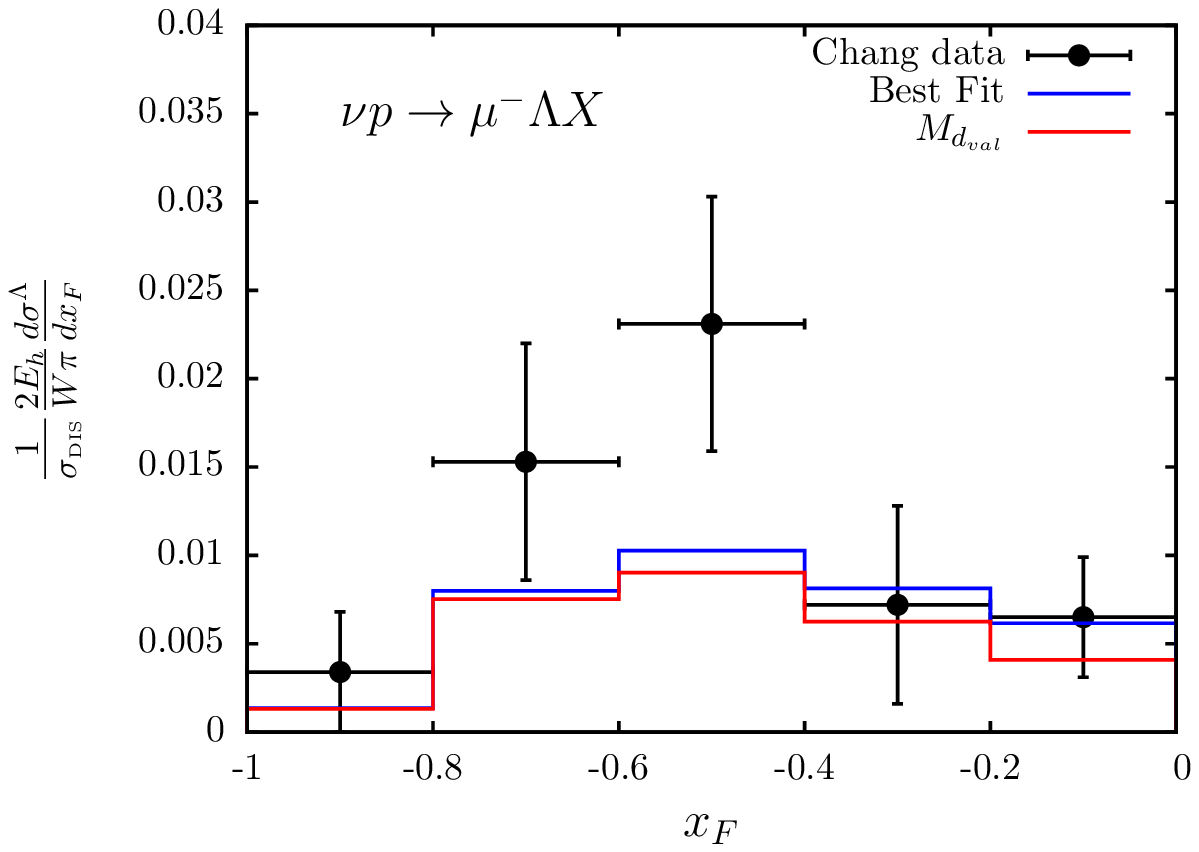}
\includegraphics[width=8cm,height=6cm,angle=0]{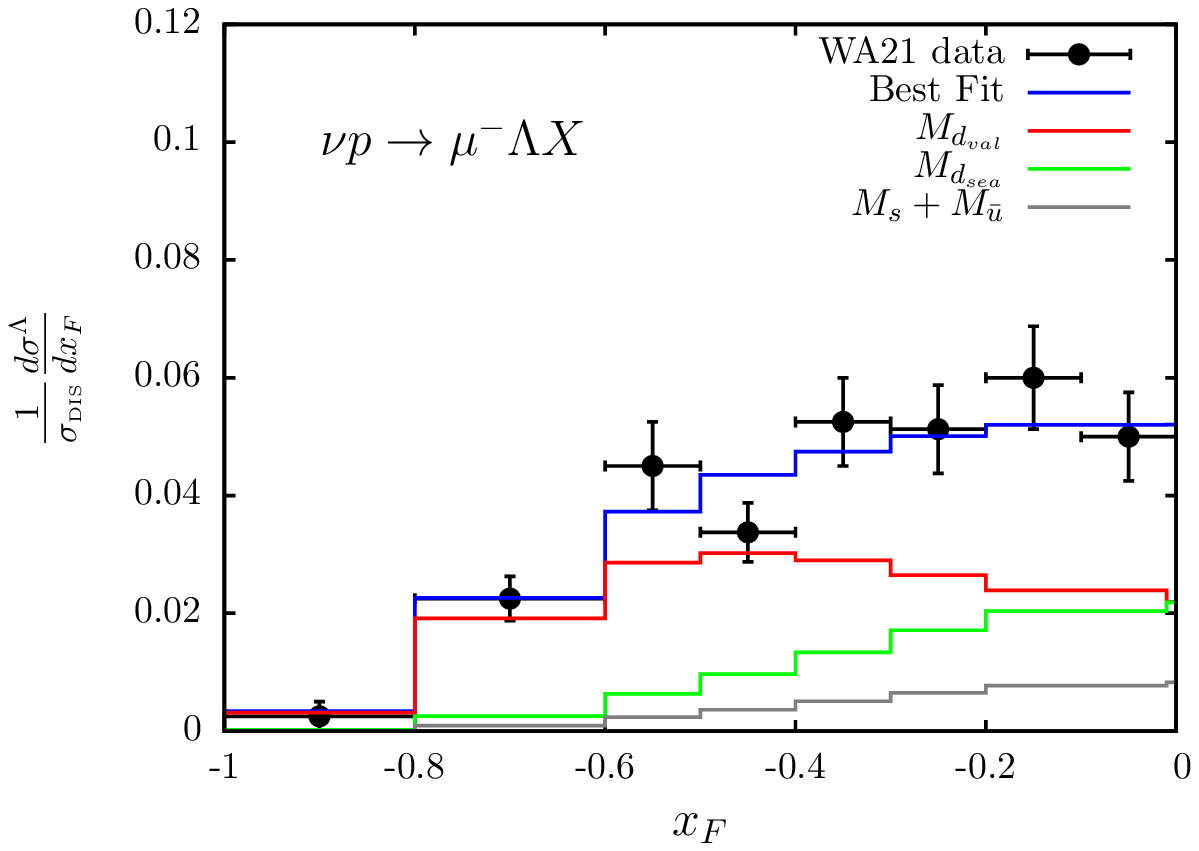}\\
\includegraphics[width=8cm,height=6cm,angle=0]{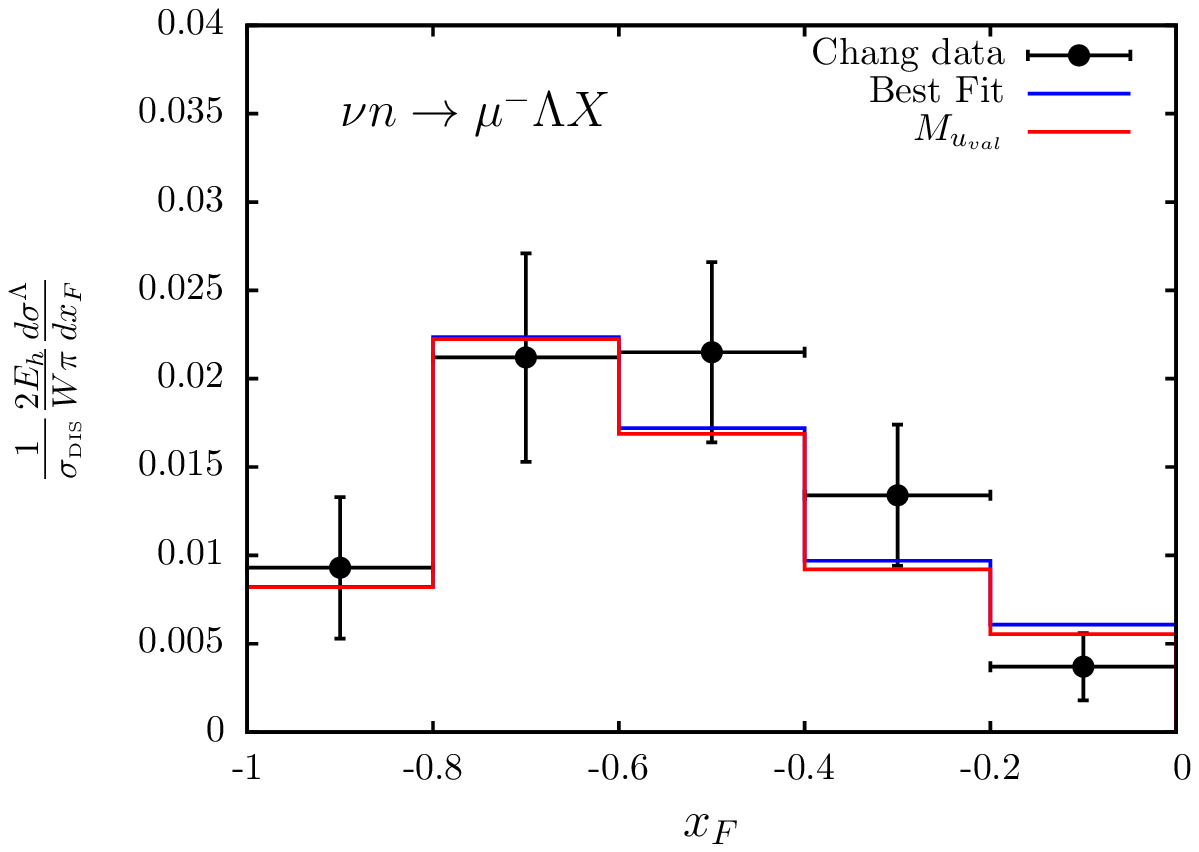}
\includegraphics[width=8cm,height=6cm,angle=0]{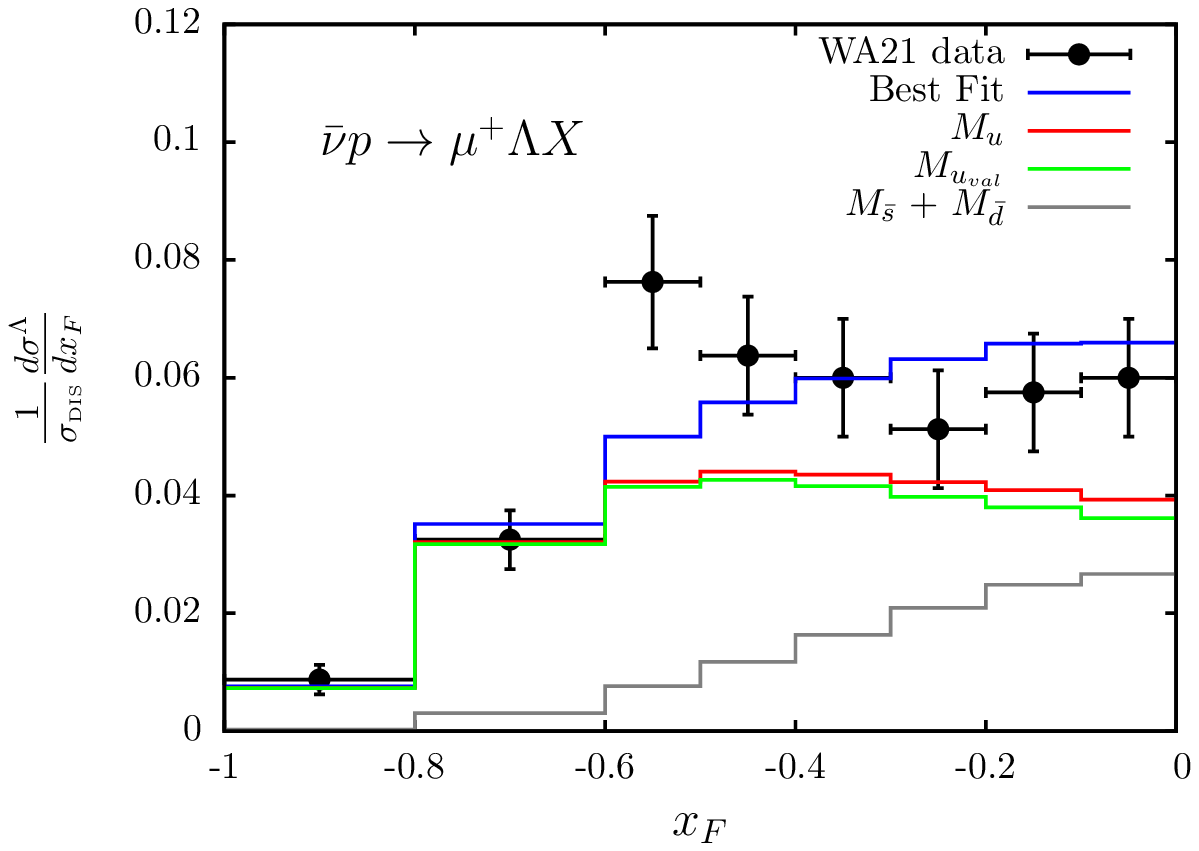}
\caption{  \small{Best-fit predictions compared to 
normalised $x_F$ distributions for charged current semi-inclusive Lambda cross-sections from 
Ref.~\cite{Chang} (left panels) and Ref.~\cite{WA21} (right panels). Various quark-flavour proton-to-Lambda fracture functions contributions are shown. Note the additional factor 
$2E_h/(\pi \, W)$ which multiplies the normalised cross-sections from Ref.~\cite{Chang}.}}
\label{Fig1}
\end{center}
\end{figure} 
 
The best-determined parameters are the three normalisations, which implies that 
three independent distributions are in fact sufficient to handle the normalisation spread between 
the various data sets.   
The four parameters controlling the shape of the spectator functions have substantially larger uncertainties.
The sources of the latter are primarily related to the intrinsic correlation between the $\alpha_i$ and
$\beta_i$ parameters introduced by the specific functional form assumed for 
the spectator functions in eq.~(\ref{zinput}). We also note that, 
at low centre-of-mass energy, where mass corrections introduced via eq.~(\ref{variables3})
are sizeable, a large portion of the $x_F$ spectrum is controlled by the behaviour 
of the spectator functions in the neighbourhood of $z=1$; the shape of the
predicted spectrum is thus mainly determined by the $\beta_i$ parameter
alone. Therefore, parametrisations with more modulable behaviour at large $z$ could improve
the description of the $x_F$ spectra. These improvements are probably marginal given the quality 
of the data used in this fit, but they might be necessary when dealing with 
higher quality data as presented, for example, in Ref.~\cite{NOMAD}. 
\begin{figure}[t]
\begin{center}
\includegraphics[width=8cm,height=6cm,angle=0]{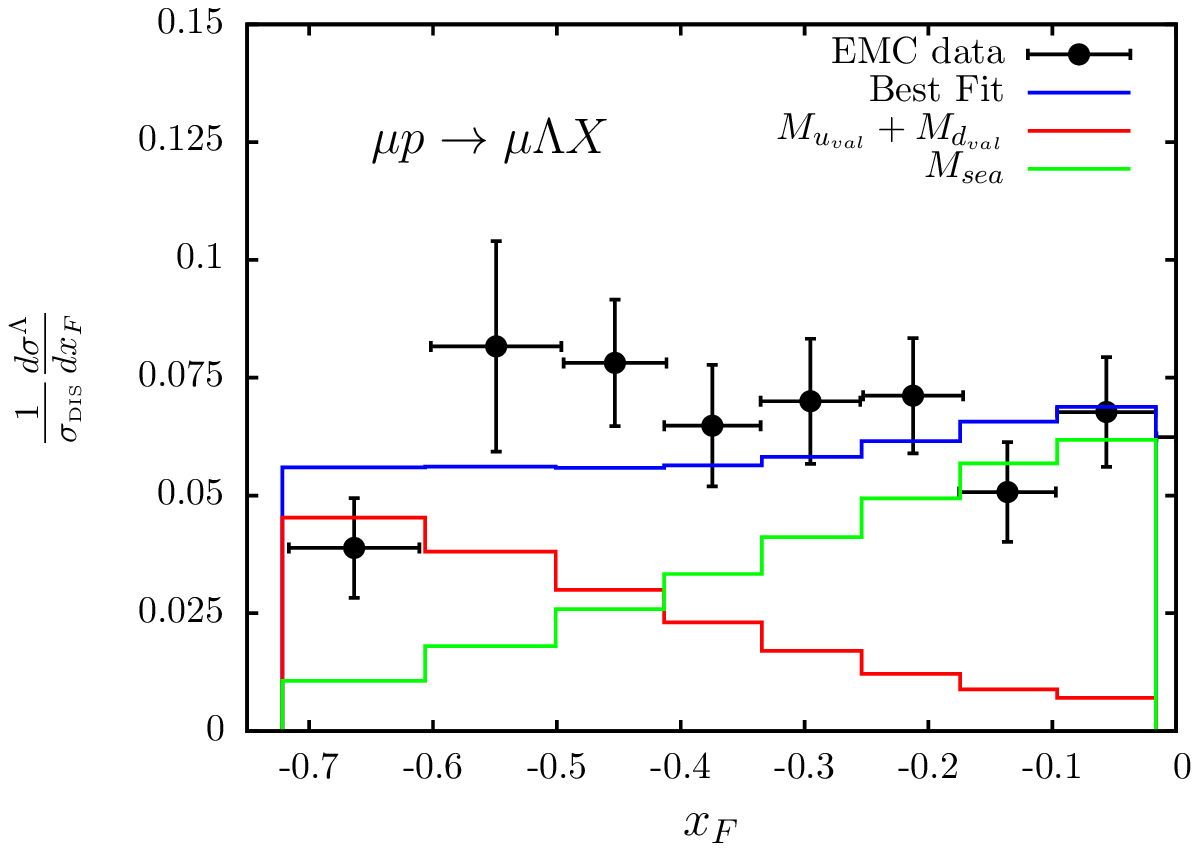}
\includegraphics[width=8cm,height=6cm,angle=0]{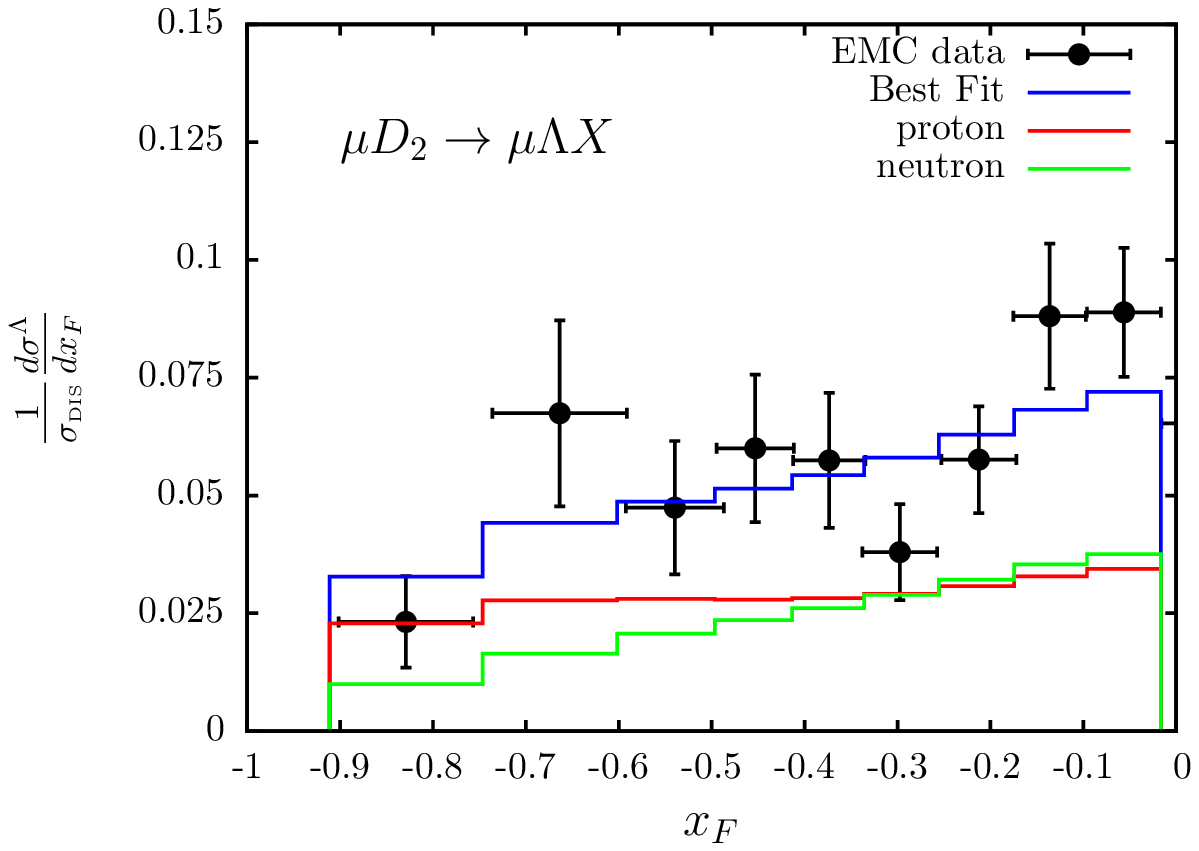}
\includegraphics[width=8cm,height=6cm,angle=0]{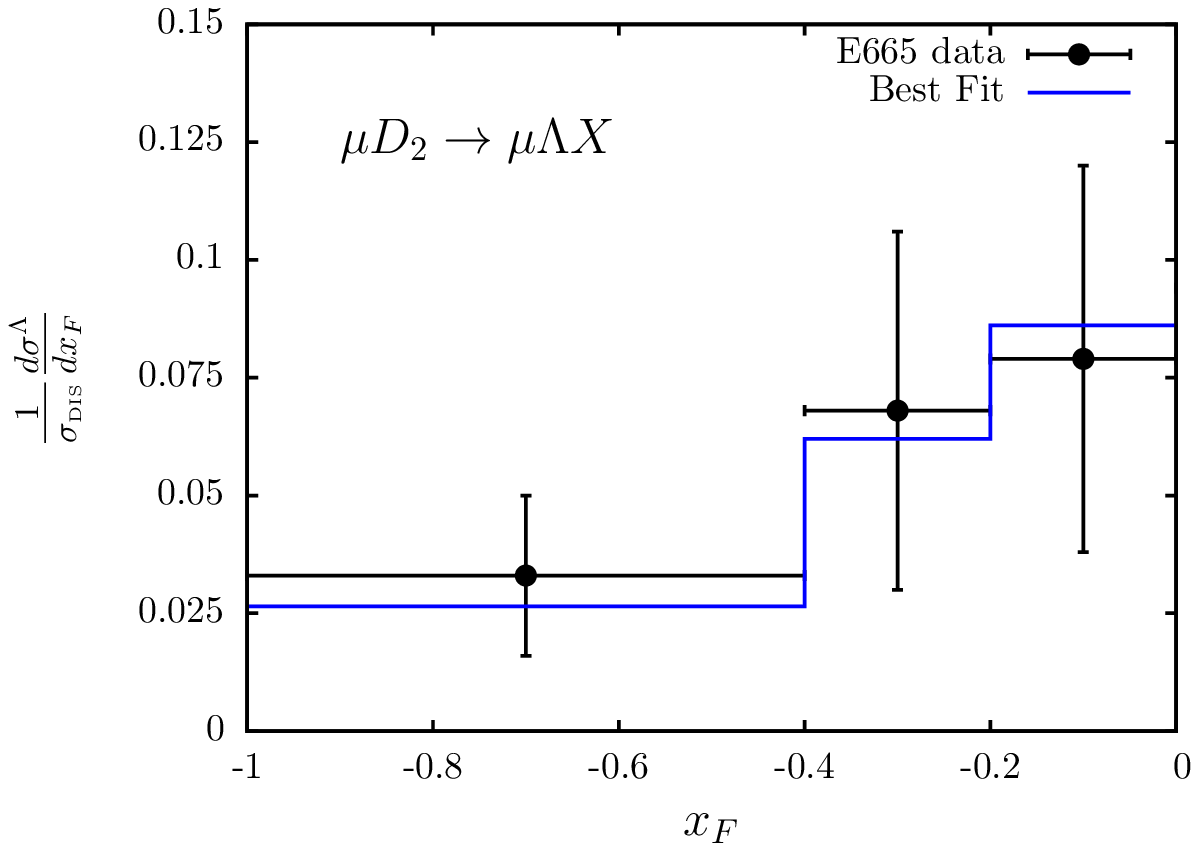}
\caption{  \small{Best fit predictions compared to normalised $x_F$ distributions for $\mu p$ (left panel) and  
$\mu D_2$ (right panel) data from Ref.~\cite{EMC} and $\mu D_2$ (bottom panel) data from 
 Ref.~\cite{E665}. In the  $\mu p$ case valence- and sea-quark fracture functions contributions are separately shown. 
In the $\mu D_2$ case the proton- and neutron-target contributions are separately shown.}}
\label{Fig2}
\end{center}
\end{figure} 

Given the stringent cut on the DIS selection, the data of Ref.~\cite{Chang} constrain the valence-quark fracture-function contributions, $M_{q_{val}}$, which in fact almost saturate the spectrum, as shown in the left column plots of Fig.~(\ref{Fig1}).
The plots in the first row of Fig.~(\ref{Fig1}) show instead a normalisation 
tension between $\nu p$ data from Ref.~\cite{Chang} and Ref.~\cite{WA21}
which, however, can be tolerated in view of the partial $\chi^2$ presented in Tab.~(\ref{bestfit}). The plots in the second row show a slight shape deformation 
in the predictions for $\nu n$ data from Ref.~\cite{Chang} which is probably due to the normalisation constraint induced by $\bar{\nu}p$ data from Ref.~\cite{WA21} on the individual $M_{u_{val}}$ and $M_{q_{sea}}$ distributions.
These results indeed indicate, as expected, that Lambdas are produced more abundantly and more forward by the fragmentation of $ud$-spectator system with respect to a $uu$-one. 

We show in Fig.~(\ref{Fig2}) the best-fit predictions for data from Ref.~\cite{EMC} and Ref.~\cite{E665}
for which the incident muon beam energy is significantly higher than neutrino and anti-neutrino 
ones. Quite interestingly, the spectrum at large $|x_F|$ is dominated by valence-quark fracture functions, 
as shown in the upper left panel of Fig.~(\ref{Fig3}). 
The differences between $M_{u_{val}}$ and $M_{d_{val}}$ distributions is
responsible for the different 
large $|x_F|$ behaviour on different targets, as shown in the upper right panel of Fig.~(\ref{Fig3}). 
In these processes, characterised by higher values of $\langle W^2 \rangle$, 
Lambda-mass corrections play a less prominent role and therefore the $z$ shape of the sea-quark fracture functions, 
as parametrised in Tab.~(\ref{bestfit}), are clearly visible in the plots. 
The quality of the fit is stable against variations of the arbitrary scale $Q_0^2$
at which fracture functions are factorised into parton distributions and spectator fragmentation functions.
The $\chi^2$ function, in fact, shows a very mild dependence on $Q_0^2$ as the latter 
is varied below the measured range between 0.5 and 1.0~GeV$^2$.

\begin{figure}[t]
\begin{center}
\includegraphics[width=8cm,height=6cm,angle=0]{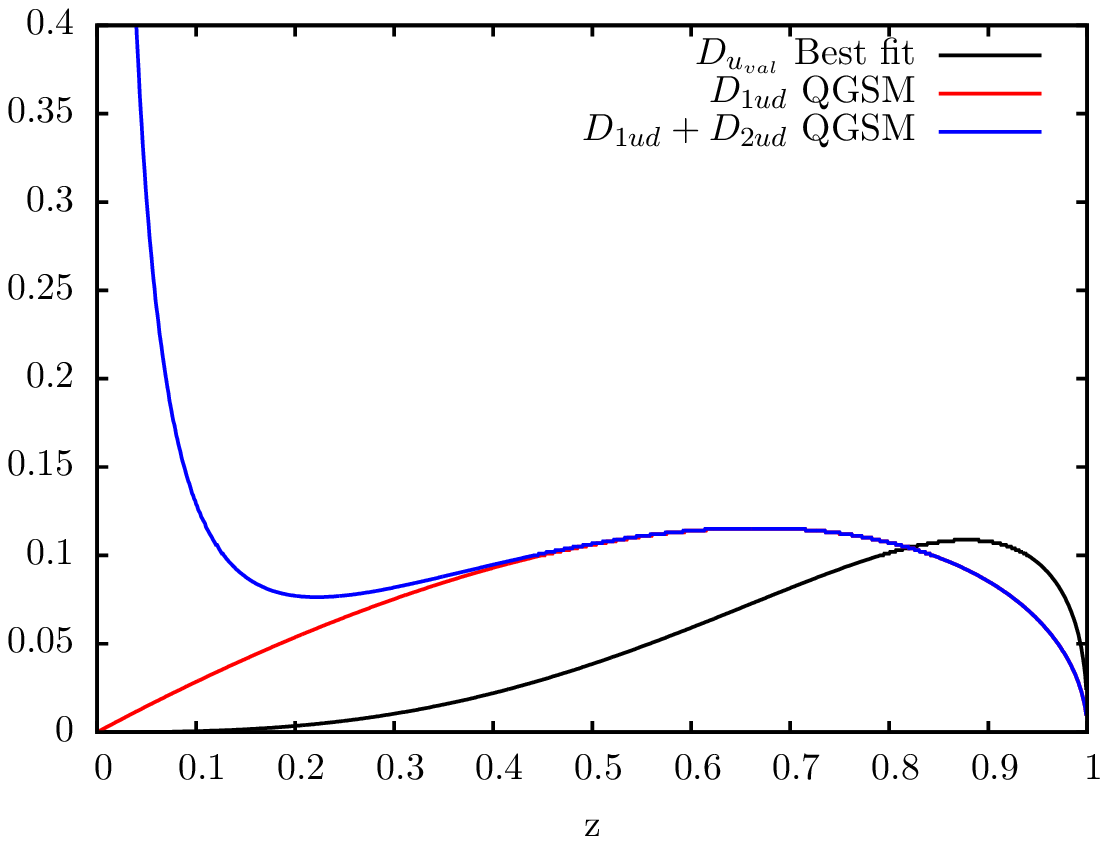}
\includegraphics[width=8cm,height=6cm,angle=0]{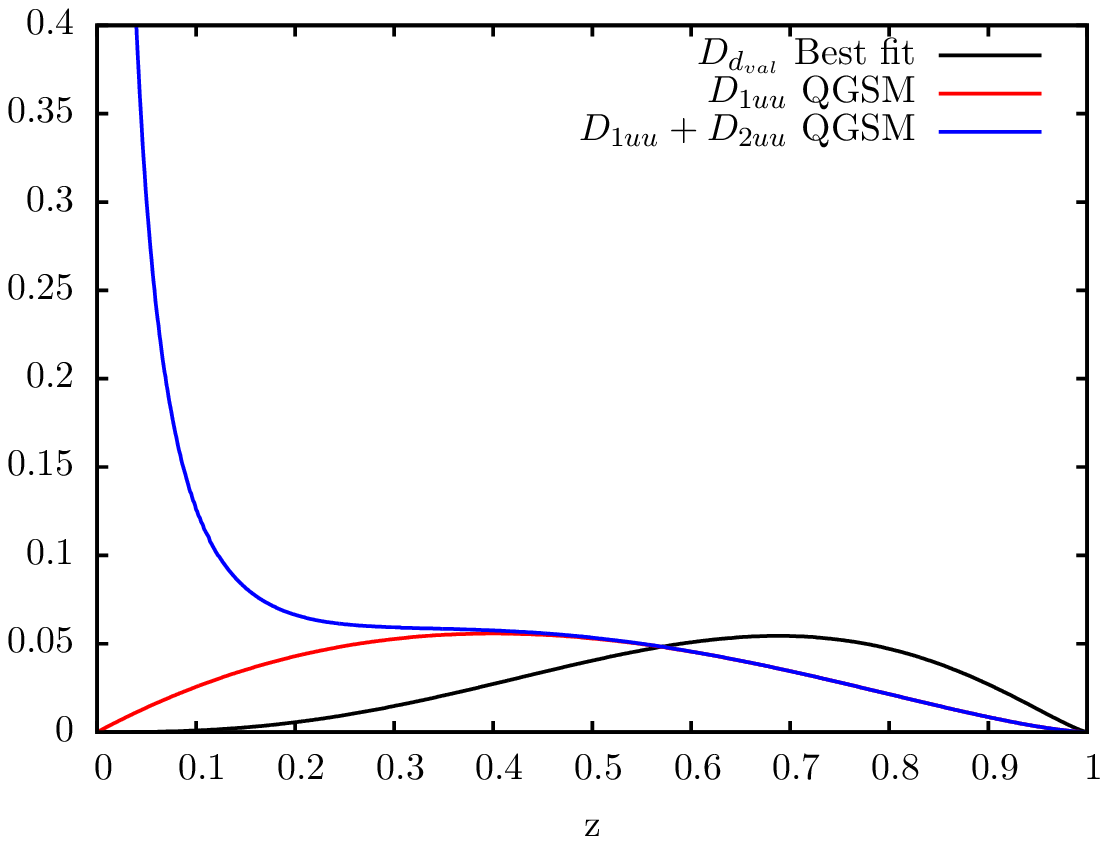}
\includegraphics[width=8cm,height=6cm,angle=0]{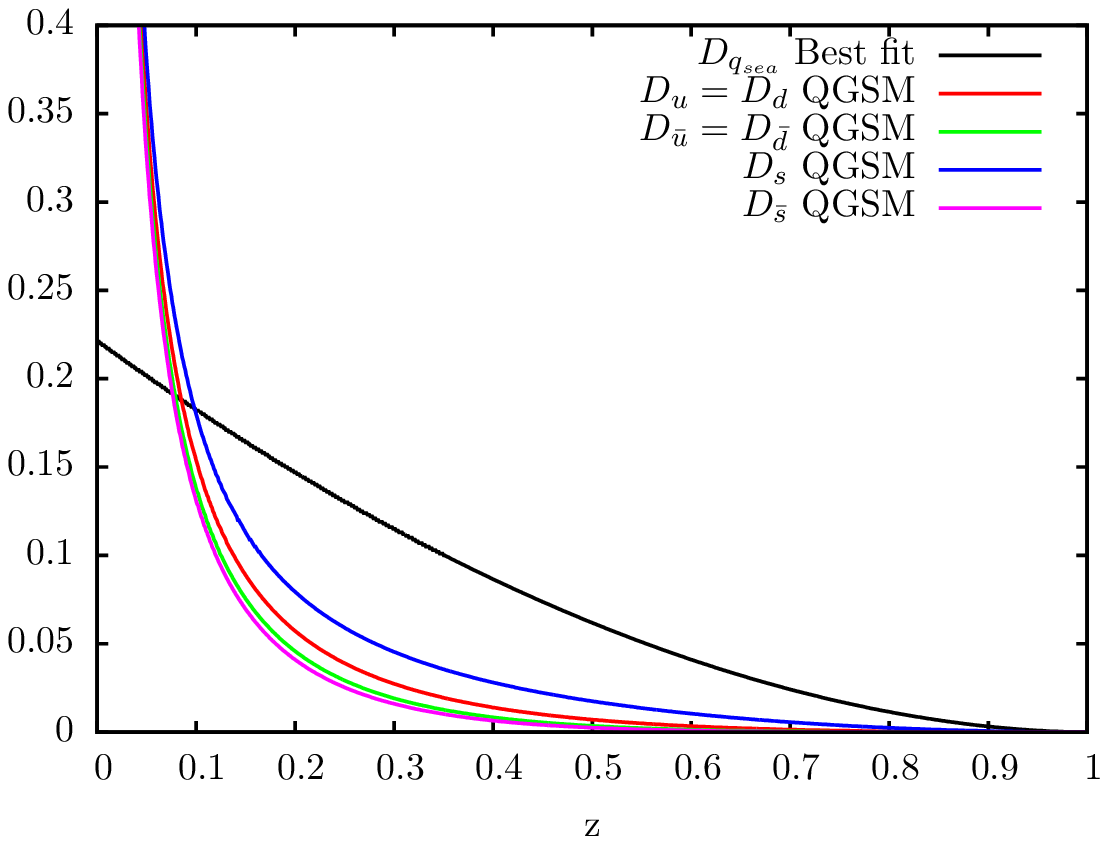}
\caption{\small{Spectator fragmentation functions from 
our best fit and from the QGS model.}}
\label{Figa}
\end{center}
\end{figure}
%
%
We wish to conclude this Section by comparing our best-fit parametrisations 
for the spectator-fragmentation functions with the analogous ones used 
in the QGS model~\cite{QGSM}. This is possible since both models assume a factorised
input at some low scale, eq.~(\ref{inputFF}), and similar 
functional forms for $\widetilde{D}_i^{\Lambda/p}(z)$, eq.~(\ref{zinput}).
The comparison however should be performed with some care:
valence- and sea-quarks distributions in the QGSM
have all the same $x$ distribution, at variance with the ones used in the present 
analysis~\cite{GRVproton}. Furthermore the scale at which the QGSM model is
assumed to be valid does not necessarily match the $Q_0^2$ scale used in the fit. This is particularly important since the scale $Q_0^2$ determines 
the relative weight of sea- and valence-parton distributions appearing 
in the fracture function decomposition in eq.~(\ref{ic1})
and therefore the $\widetilde{D}_i^{\Lambda/p}(z)$ themselves.
Comparisons for the fragmentation functions are presented in Fig.~(\ref{Figa}).
The Lambda spectrum at large $|x_F|$ is driven by 
the large-$z$ behaviour of $M_{u_{val}}$ and $M_{d_{val}}$.
We can then compare directly the best-fit $\beta_{i}$ parameters 
with the analogous ones in the QGS model. 
They can be read out from the $D_{1ud}^{\Lambda}$ 
and $D_{1uu}^{\Lambda}$ terms appearing in the appendix of
Ref.~\cite{QGSM} and, in our notation, are given 
by $\beta_{u_{val}}=0.5$ and $\beta_{d_{val}}=1.5$.
These values, within errors, are in agreement with the best-fit parameters
extracted in this analysis and listed in Tab.~(\ref{bestfit}). 
The behaviour of the $D_{1ud}^{\Lambda}$ and $D_{1uu}^{\Lambda}$ terms  
at low $z$ is significantly softer, $\alpha_{q_{val}}=1$, than in our model, 
$\alpha_{q_{val}}=2.82$. 
In our model, on the other hand, sea-quarks spectator fragmentation-functions
are harder, $\beta_{q_{sea}}=1.8$ and $\alpha_{q_{sea}}=0$,
with respect to the QGSM predictions, for which 
$\beta_{q_{sea}}=2\div3.5$ depending on the sea-quarks flavour 
and $\alpha_{q_{sea}}=-1$. 
The main differences between the two models therefore appear at intermediate and low $z$
where there is no evidence in our best-fit parametrisations of the 
$z^{-1}$ behaviour as predicted by the QGSM.
Such different behaviour is determined by the value 
of the $\alpha_{q_{sea}}$ parameter which, 
due to the reduced sensitivity at low $z$ caused 
by sizeable Lambda-mass corrections, was loosely determined by the fit 
and therefore fixed to zero. With these respect semi-inclusive DIS data
at higher centre-of-mass energy will better constrain the spectator
fragmentation functions at small $z$ and test QGSM predictions.


\section{Predictions}
\label{predictions}
\begin{figure}[t]
\begin{center}

\includegraphics[width=8cm,height=6cm,angle=0]{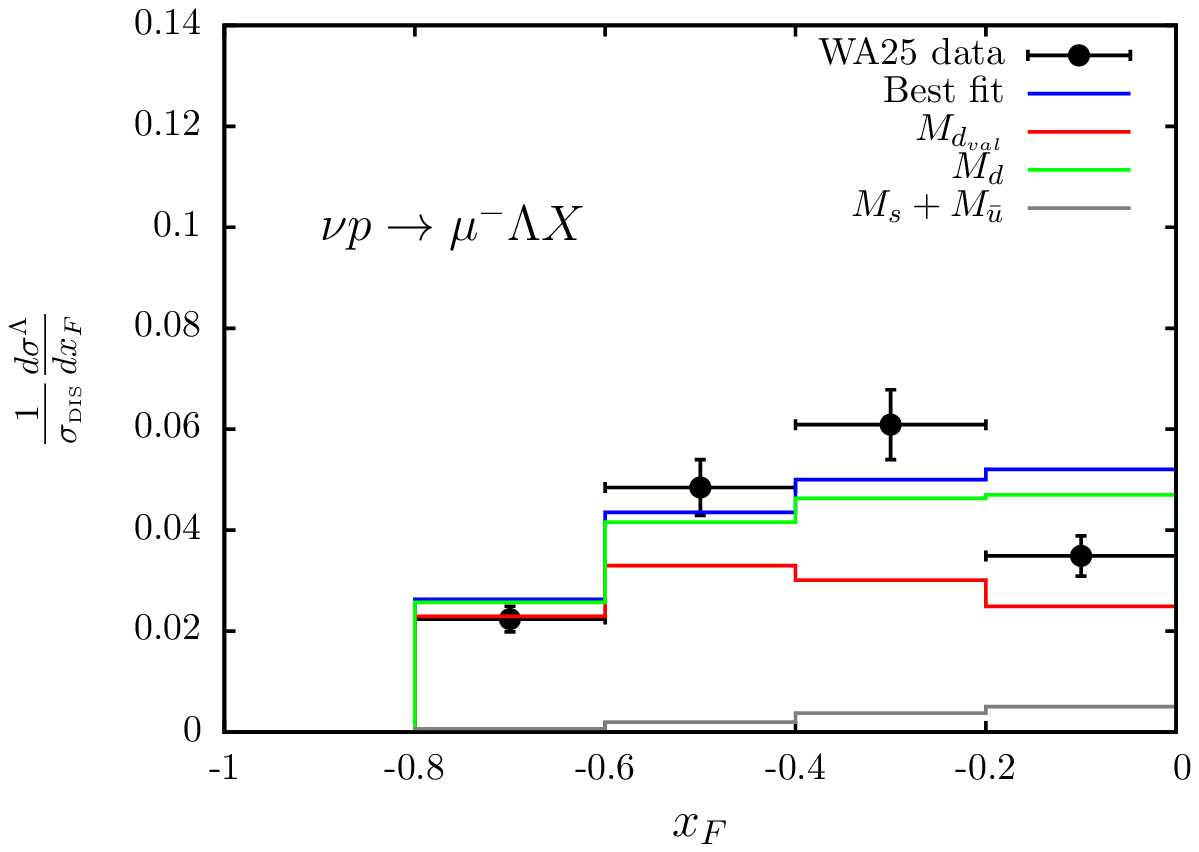}
\includegraphics[width=8cm,height=6cm,angle=0]{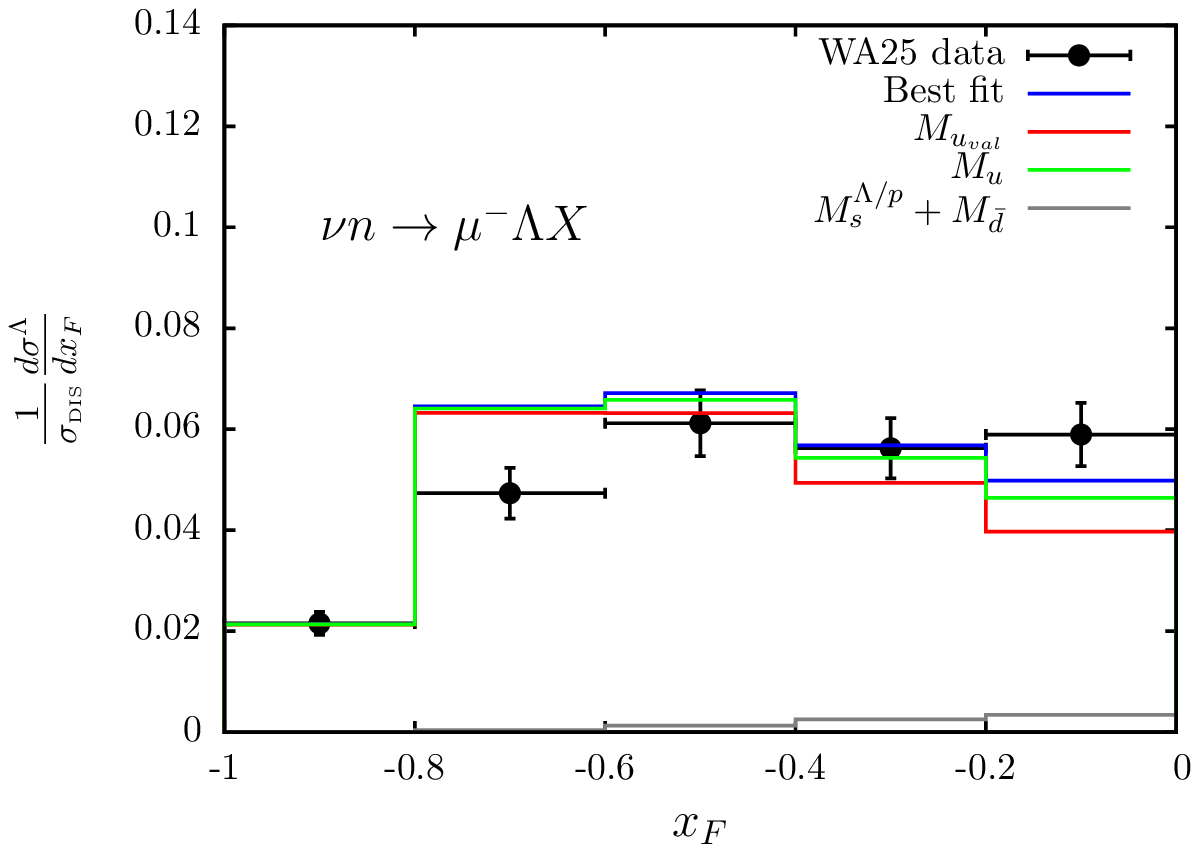}\\
\includegraphics[width=8cm,height=6cm,angle=0]{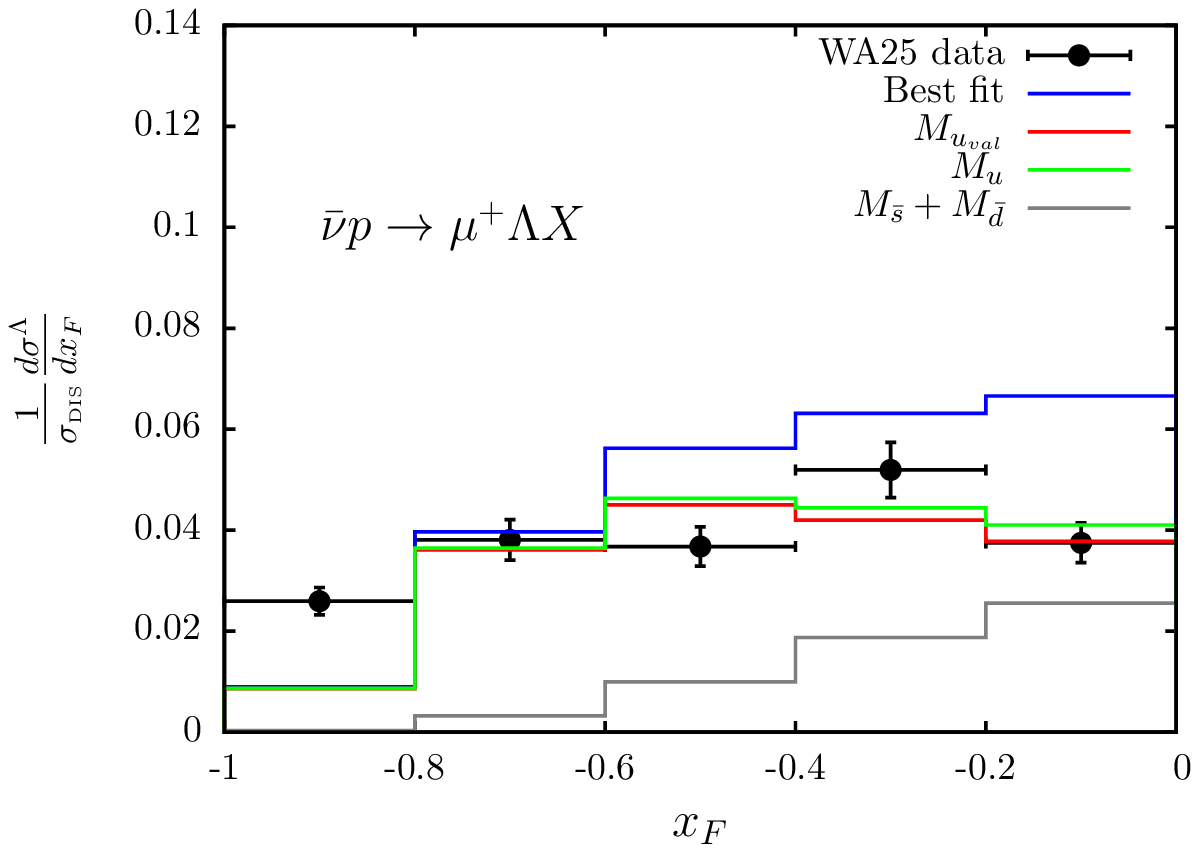}
\includegraphics[width=8cm,height=6cm,angle=0]{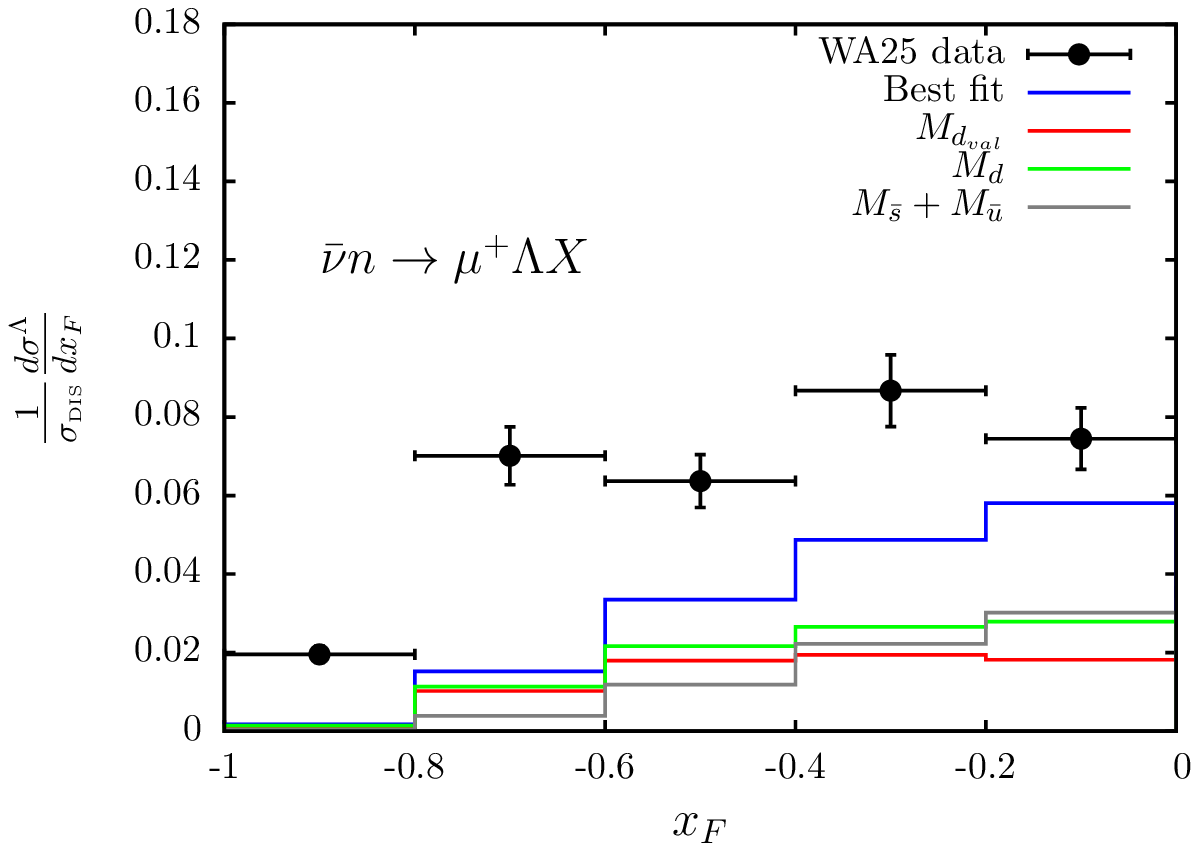}
\caption{  \small{Best-fit predictions compared to 
normalised $x_F$-distributions for charged current semi-inclusive Lambda cross-sections from 
Refs.~\cite{Allasia,Allasia2}. Various quark-flavour proton-to-Lambda fracture functions ontributions are shown.}}
\label{Fig3}
\end{center}
\end{figure} 

In this last section we wish to discuss the degree of predictivity of the model. 
It is of particular importance to determine to which extent the model is able to 
reproduce $x_F$ distributions coming from data not included in the fit or  
observables other than the ones used in the fit.
This comparison will eventually pinpoint sectors of the model which 
may need to be improved. 
The comparison with the $x_B$ or $Q^2$ distributions
require the reconstruction 
of the current fragmentation term, which is obtained with the semi-inclusive version of 
eqs.~(\ref{LOeP},\ref{LOnuP},\ref{LOnubarP}) where fracture functions are substituted by appropriate products of parton distributions and fragmentation functions. Such term has been estimated to leading order by using the parton distributions of Ref.~\cite{GRVproton} and
the fragmentation functions of Ref.~\cite{AKK}.

As a first example, we consider Lambda production in neutrino- and anti-neutrino-induced 
charged-current DIS on proton and neutron targets~\cite{Allasia}. These data 
provide full flavour discrimination of the spectator system and can be used 
to test the assumptions made in eq.~(\ref{isospin}). 
Unfortunately, the $x_F$ distributions for these data are presented in Ref.~\cite{Allasia}
as histograms without errors. In a subsequent paper by the same collaboration~\cite{Allasia2},
the Lambda yields are updated but no $x_F$ distributions are given.

In order to gauge the agreement of the best-fit model predictions with these data, 
we have assessed the errors in the following way. 
First the histograms from Ref.~\cite{Allasia} have been scaled down to the updated yields of Ref.~\cite{Allasia2} assuming no change in their shape. Then we have assumed that the relative errors of the $x_F$ distributions in each $x_F$ bin are equal. This in turn implies 
that they are respectively equal 
to the relative error on the corresponding Lambda yield. This procedure guarantees that upon 
integration over $x_F$, the experimental yield and its error are correctly
recovered.
The best-fit predictions are compared to data in Fig.~(\ref{Fig3}).
We note that, in general, the cross sections on proton target are fairly reproduced in normalisation, as expected, since the Lambda yields in $\bar{\nu} p$- and $\nu p$-scattering of Ref.~\cite{Allasia2} are 
in agreement within errors with those of data included in the fit~\cite{Chang,WA21}.
The predicted distributions do not reproduce the turn-over at low $|x_F|$ as seen in data,  reflecting the shape of the fitted data~\cite{Chang,WA21},
as shown in the right-hand-side panels of Fig.~(\ref{Fig1}).
The comparison with neutron-target data reveals that, in this case, the model does not perform 
equally well. In particular, the $\nu n$ cross sections peaks at too 
large values  of $|x_F|$, a behaviour mainly driven by the $\nu n$ data of Ref.~\cite{Chang} 
and illustrated in the lower left panel of Fig.~(\ref{Fig1}).
While all these distributions are quite well reproduced in normalisation, 
the model significantly underestimates the $\bar{\nu} n$-scattering 
data. In this case, the disagreement between data and theory might indicate that
some of the assumptions on fracture-function parametrisations used in the fit, for example 
$\widetilde{D}^{\Lambda/n}_{u(dd)}=\widetilde{D}^{\Lambda/p}_{d(uu)}$
as well as considering a common spectator fragmentation function
for all sea-quarks, could be relaxed, as we discuss in the following.

\begin{figure}[t]
\begin{center}
\includegraphics[width=8cm,height=6cm,angle=0]{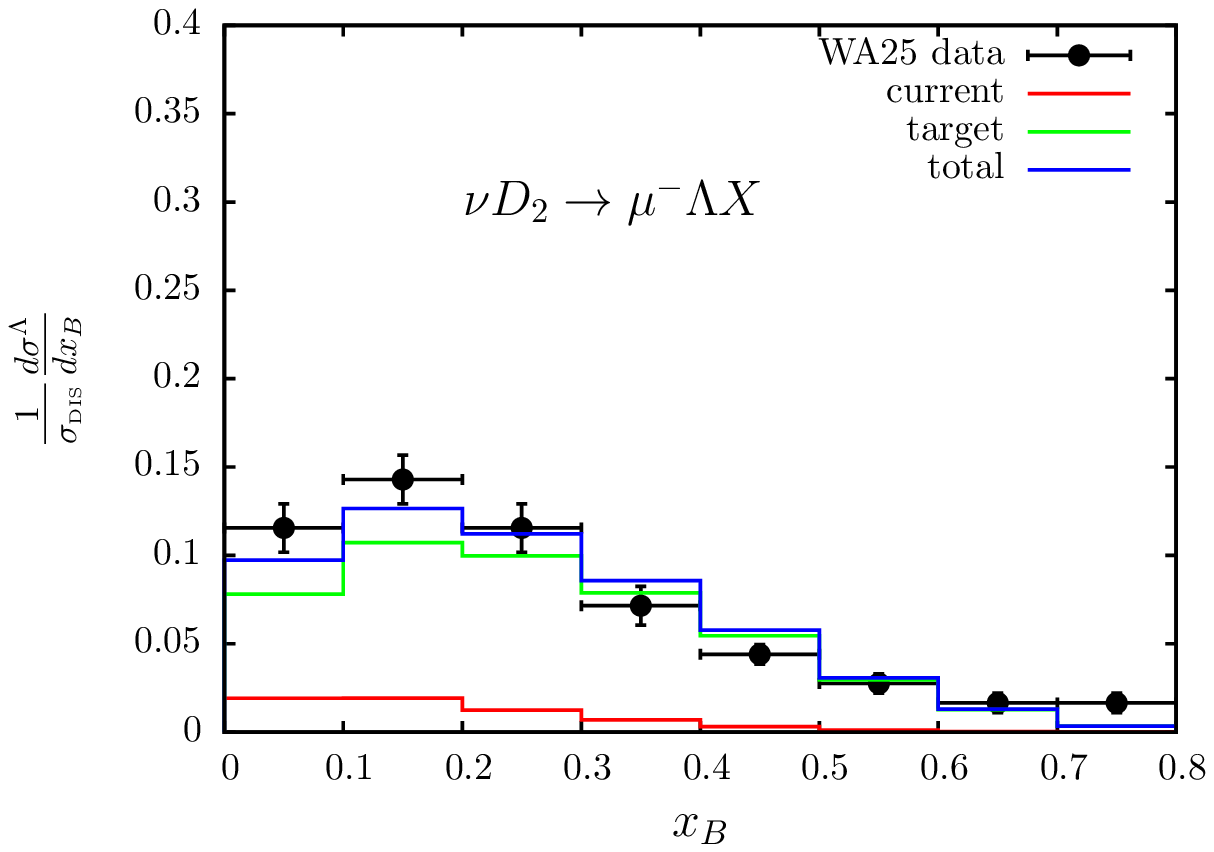}
\includegraphics[width=8cm,height=6cm,angle=0]{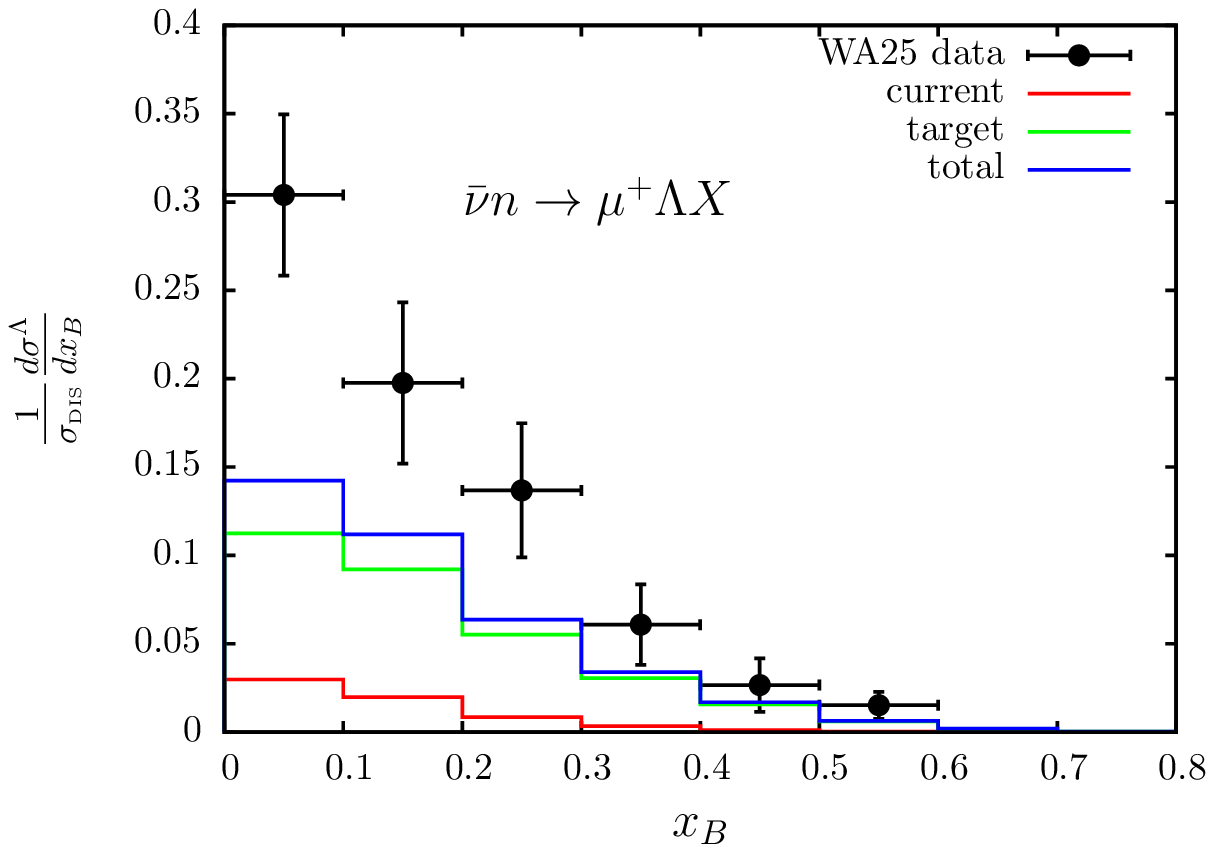}\\
\caption{\small{Normalised $x_B$ distributions in $\nu D_2$ (left panel) 
and $\bar{\nu} n$ (right panel) scattering. Data are taken from Ref.~\cite{Allasia2}.}}
\label{Fig4}
\end{center}
\end{figure}
The same data are presented in Fig.~(\ref{Fig4}) in terms of normalised $x_B$ distributions~\cite{Allasia2}. 
We remind the reader that we integrate over the $x_B$ variable in the
reconstruction of the $x_F$ distributions.
Therefore the comparison of the $x_B$ distributions with data constitutes a non-trivial test for the model.  
Excellent agreement is found for the $\nu D_2$ cross section, both in shape and in normalisation.
On the other hand, model predictions for the $\bar{\nu} n$ cross section fail to
reproduce experimental distributions, as in the corresponding $x_F$ distribution in the lower right panel of Fig.~(\ref{Fig3}).   
Since the $\bar{\nu} n$ cross section significantly exceeds the $\nu D_2$ only in the smallest $x_B$ bin, the hypothesis that target Lambdas could be more abundantly produced in $\bar{\nu} \bar{s}$-scattering rather
than in $\nu s$-scattering was originally formulated in Ref.~\cite{Allasia2},
based on the idea that, in the sea-quark region, $\bar{\nu}
\bar{s}$-scattering leaves the correct hypercharge in the target
spectator. In our framework, this hypothesis would directly translate into an asymmetry of the spectator fragmentation functions in the strange sector, $\widetilde{D}_{\bar{s}}^{\Lambda/p} \neq \widetilde{D}_s^{\Lambda/p}$. 
The cross sections used in the regression however can not individually constrain 
$\widetilde{D}_{\bar{s}}^{\Lambda/p}$ and $\widetilde{D}_s^{\Lambda/p}$,
because of the linear combinations of sea-quark fracture functions 
appearing in the structure functions in eqs.~(\ref{LOeP},\ref{LOnuP},\ref{LOnubarP}).

In this respect, the associated production of target Lambdas in dimuon (anti-)neutrino-nucleon deep-inelastic scattering, $\nu N \rightarrow \mu^- \mu^+  \Lambda X$ and 
$\bar{\nu} N \rightarrow \mu^+ \mu^-  \Lambda X$,
could represent an alternative sensitive test of the latter hypothesis.
In the corresponding parton sub-process, $W^+ s\rightarrow c$
and $W^- \bar{s}\rightarrow \bar{c}$, a (anti-)charm, produced by a charged current, 
semileptonically decays in the current fragmentation region into a final-state, secondary, muon. 
Given the small off-diagonal quark-mixing CKM matrix elements and the known 
parameters of charm quarks decay, the detection of the additional muon  
allows to probe directly the (anti-)strange component of the nucleon~\cite{petti}.
Therefore, in this class of events,  the additional detection of a $\Lambda$ hyperon in the target fragmentation could test the the proposed asymmetry.

\begin{figure}[t]
\begin{center}
\includegraphics[width=8cm,height=6cm,angle=0]{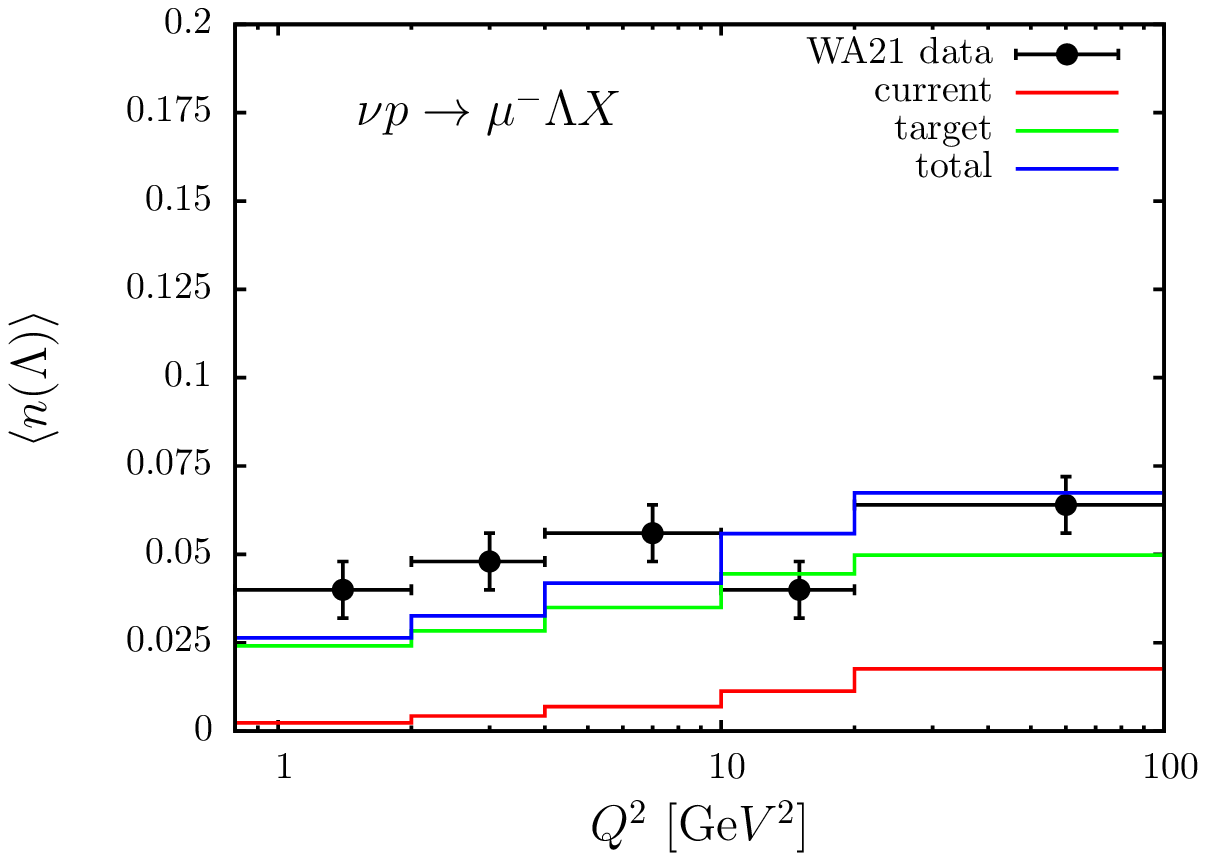}
\includegraphics[width=8cm,height=6cm,angle=0]{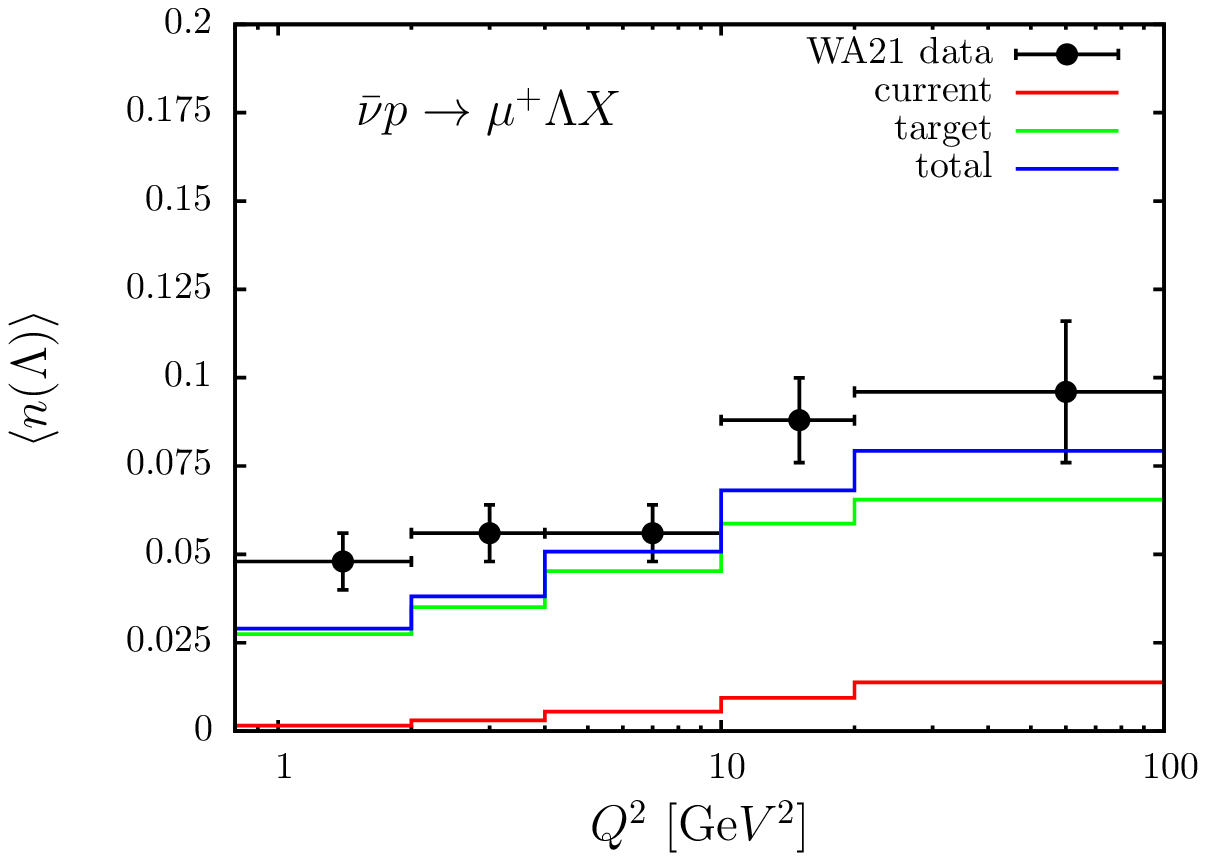}
\caption{\small{Lambda average multiplicity as a function of $Q^2$ in $\nu p$ (left panel) 
and $\bar{\nu} p$ (right panel) scattering. Current and target contributions are separately shown. Data are from Ref.~\cite{WA21}.}}
\label{Fig5}
\end{center}
\end{figure}

We conclude by showing in Fig.~(\ref{Fig5}) the model predictions for 
the averaged Lambda multiplicity as a function of $Q^2$ in $\nu p$ and $\bar{\nu} p$ charged-current SIDIS cross sections.
Although the $x_F$ distributions from this data 
set~\cite{WA21} have been already used in the fit and therefore agreement in normalisation might be expected, 
the model is able to reproduce reasonably well the observed $Q^2$ dependence, 
with a tendency to undershoot the data at the lowest value of $Q^2$.
Since the $Q^2$ dependence built in the model via fracture-function evolution equations 
can be considered one of the most stringent predictions of the underlying theoretical 
framework, the reasonable agreement between data and predictions
can be considered the first step towards a conclusive validation of the perturbative framework.
The validation procedure and the model itself would, in fact, highly benefit from the constraints imposed by multi-differential 
distributions. Cross sections at fixed $x_B$ and $x_F$ as a function of $Q^2$, for example, could give access, through scaling violations, to the presently unconstrained
gluon fracture function, $M_g^{\Lambda/p}(x_B,z,Q^2)$.  

\section{Summary and conclusions}
\label{Conclusions} 
In this paper we have analysed experimental data on the production of
Lambda hyperons in the SIDIS target fragmentation region in terms of fracture functions. 
A model for the latter has been proposed and the free parameters 
appearing in the input distributions have been fixed by performing a fit 
to a variety of neutral- and charged-current semi-inclusive DIS cross sections. 
The main features seen in the data can be fairly reproduced by the model. 
In particular the spectator-fragmentation functions associated with 
the removal of valence quarks populate the very forward part of the 
$x_F$-spectrum at large and negative values of $x_F$. 
On the other hand, the sea-quarks contribution is concentrated at small 
and negative values of $x_F$. The predictions based on the model 
are in fair agreement with data not included in the fit and with observables 
depending on variables which are integrated over in the analysis, 
especially the $Q^2$ dependence of the Lambda multiplicity, which 
is a stringent test of the underlying theoretical framework. 
Although further tests and additional experimental informations are necessary to validate and eventually improve the model, 
it may be used to quantitatively investigate   
spectator-fragmentation mechanisms within a perturbative QCD approach. Since higher-order calculations 
are available in the literature, the analysis can be extended to next-to-leading order accuracy.
The model can be easily generalised to take into account Lambda polarisation allowing 
spin-transfer studies~\cite{ellis} in the target region and it may  
find application in the estimation of nuclear corrections to target fragmentation. 

\section*{Acknowledgements}

We gratefully acknowledge M.~Stratmann and S.~Albino for providing us with their fragmentation-function
routines. We especially acknowledge D.~Naumov for interesting discussions related to backgrounds
in Lambda production in DIS and for providing us the neutrino flux parametrisations. 
We wish to thank the organizers of the Workshop "Strangeness polarization in semi-inclusive and exclusive Lambda production" held in ECT*, Trento, in October 2008 and all the participants for stimulating discussions on this topic. 
We finally thank Laurent Favart, Dmitry Naumov, Jean-Ren\'e Cudell and Luca Trentadue for a critical reading of the manuscript prior to submission.

\end{document}